\documentclass[aps,twocolumn,prb,showpacs,floatfix]{revtex4}

\usepackage{epsfig,amsmath,amssymb}
\usepackage{graphics} 
\usepackage{subfigure}
\usepackage{color}
\usepackage{multirow}

\begin{document}

\title
{
Spin-1/2 Heisenberg antiferromagnet on an anisotropic kagome lattice
}
\author
{P.~H.~Y.~Li and R.~F.~Bishop}
\affiliation
{School of Physics and Astronomy, Schuster Building, The University of Manchester, Manchester, M13 9PL, UK}

\author
{C.~E.~Campbell}
\affiliation
{School of Physics and Astronomy, University of Minnesota, 116 Church Street SE, Minneapolis, Minnesota 55455, USA}

\author
{D.~J.~J.~Farnell}
\affiliation 
{Division of Mathematics and Statistics, Faculty of Advanced Technology, 
University of Glamorgan, Pontypridd CF37 1DL, Wales, UK}

\author
{O. G\"otze and J. Richter}
\affiliation
{Institut f\"ur Theoretische Physik, Otto-von-Guericke Universit\"at Magdeburg, 39016 Magdeburg, Germany}

\begin{abstract}

We use the coupled cluster method to study the zero-temperature properties of an
extended two-dimensional Heisenberg antiferromagnet formed from spin-1/2 moments
on an infinite spatially anisotropic kagome lattice of corner-sharing isosceles
triangles, with nearest-neighbor bonds only.  The bonds have exchange constants
$J_{1}>0$ along two of the three lattice directions and $J_{2} \equiv \kappa J_{1} > 0$
along the third.  In the classical limit the ground-state (GS) phase for 
$\kappa < 1/2$ has collinear ferrimagnetic (N\'{e}el$'$) order where the $J_2$-coupled 
chain spins are ferromagnetically ordered in one direction with the remaining spins
aligned in the opposite direction, while for $\kappa > 1/2$ there exists an infinite
GS family of canted ferrimagnetic spin states, which are energetically degenerate.
For the spin-1/2 case we find that quantum analogs of both these classical states
continue to exist as stable GS phases in some regions of the anisotropy parameter
$\kappa$, namely for $0<\kappa<\kappa_{c_1}$ for the N\'{e}el$'$ state and for
(at least part of) the region $\kappa>\kappa_{c_2}$ for the canted phase.
However, they are now separated by a paramagnetic phase without either 
sort of magnetic order in the region
$\kappa_{c_1} < \kappa < \kappa_{c_2}$, which includes the isotropic kagome point
$\kappa = 1$ where the stable GS phase is now believed to be a topological 
($\mathbb{Z}_2$) spin liquid.  Our best numerical estimates are
$\kappa_{c_1} = 0.515 \pm 0.015$ and $\kappa_{c_2} = 1.82 \pm 0.03$.  

\end{abstract}
\maketitle

\section{Introduction}
\label{intro}

Low-dimensional quantum magnets, especially those defined on regular two-dimensional
(2D) lattices, have been the subject of intense study in recent years (and see,
e.g., Refs. \onlinecite{2D_magnetism_1,2D_magnetism_2} for recent reviews).  In
particular it is known that highly frustrated 2D quantum antiferromagnets display a bewilderingly
rich panoply of ground-state (GS) phases, which often have no classical counterparts.
Examples include various valence-bond crystalline and spin-liquid phases.  Among 
the parameters that determine the zero-temperature ($T=0$) phase diagram of such
systems are the dimensionality and structure (e.g., the coordination number) 
of the crystallographic lattice on whose sites the magnetic ions are situated, 
the spin quantum number $s$ of the ions, and the type and range of the magnetic bonds
between the ions that often compete for differing forms of order, thereby
leading to frustration.  We have learned too that the quantum versions of 
classical models that have massively degenerate ground states, especially
those with a nonzero ($T=0$) GS entropy, are prime candidates for systems
with novel GS phases.

Among all such candidate spin-lattice systems, therefore, those with periodic arrays 
of vertex-sharing structures, each of which is itself magnetically frustrated, occupy
a special niche.  These include the three-dimensional (3D) pyrochlore 
lattice of vertex-sharing tetrahedra and the 2D kagome lattice of corner-sharing triangles.  
Of these, the spin-1/2 Heisenberg antiferromagnet (HAF) on the 
2D kagome lattice has been the subject of intensive study in recent years.\cite{2D_magnetism_2,
Mar:1991,Harris:1992,Sach:1992,Huse:1992,Chalker:1992,Singh:1992,Chubukov:92,
Leung:1993,Asakawa:1994,Zeng:1995,Henley:1995,Lech:1997,Mila:1998,Wald:1998,
Hast:2000,Mamb:2000,Fa:2001,Bernhard:2002,Syr:2002,Nik:2003,Schmal:2004,Bud:2004,
Capponi:2004,Wang:2006,Sin:2007,Ran:2007,Misg:2007,Sin:2008,Herm:2008,Jiang:2008,
Henley:2009,Sindzingre:2009,Poilblanc:2010,Bishop:2010_KagomeSq,Eve:2010,
Yan:2011,Iqbal:2011a,Lauchli:2011,Lee:2011,Nakano:2011a,Tay:2011a,Iqbal:2011b,
Cepas:2011,Tay:2011b,Poilblanc:2011,Gotze:2011,Jiang:2012_kagome_spinLiquid,Shimokawa:2012,
Masuda:2012,Depenbrock:2012_kagome_spinLiquid}  Even after several decades of research the
nature of the GS phase of the spin-1/2 HAF on the spatially isotropic kagome lattice has
remained uncertain until very recently.  Various outcomes, ranging from states with
magnetic order to valence-bond solids or quantum spin liquids of different types,
have been proposed.  

Although many such studies agree on the finding that the GS phase lacks 
magnetic long-range order (LRO), there has remained
uncertainty over its precise character.  For example, some studies have favored a gapless critical spin
liquid of one type or another, while others have favored one or other valence-bond crystals
with an abundance of low-lying spin-singlet excited states.  Only in the last year or so has
compelling numerical evidence been provided,\cite{Yan:2011} due to advances in the
density-matrix renormalization group (DMRG) technique, that the ground state is both gapped and is without
any signal of either valence-bond or magnetic order at the largest finite-size systems that
could be studied.  In the past few months further convincing evidence has come from 
different large-scale DMRG studies\cite{Jiang:2012_kagome_spinLiquid,Depenbrock:2012_kagome_spinLiquid}
that this GS phase is indeed a topological ($\mathbb{Z}_2$) spin liquid, as we discuss further in 
Sec.~\ref{discussions_conclusions} below when we discuss our own results.
We should note, however, that despite these recent findings no final consensus
has yet been reached within the community as to whether the spin-liquid
ground state is actually gapped or gapless.

Theoretical interest in spin-1/2 kagome HAFs heightened considerably in the 
last few years with the discovery of several candidate materials for their experimental
realizations.  Chronologically, the first promising such candidate 
was the mineral herbertsmithite (also known as
Zn-paratacamite), $\gamma$-Cu$_3$Zn(OH)$_{6}$Cl$_{2}$,\cite{Shores:2005,Helton:2007,deVries:2008}
for which it has been shown that the spin-1/2 Cu$^{2+}$ ions are antiferromagnetically
coupled and lie on the vertices of well separated and
structurally undistorted kagome-lattice planes.  Although the underlying
kagome planes in herbertsmithite appear to be essentially structurally
perfect, there does appear to be an appreciable amount of antisite disorder
due to a mixing of the spin-1/2 Cu$^{2+}$ ions and the diamagnetic Zn$^{2+}$ ions between the
Cu and Zn sites.  This disorder acts to introduce a coupling between the kagome planes,
thereby effectively destroying the local 2D nature of the system.  Thus, while
herbertsmithite is structurally perfect, these impurities, together with
a spin-orbit coupling that can be modelled by a Dzyaloshinskii-Moriya
interaction with a non-negligible strength parameter, act to
complicate the comparison of theory with experiment.  As a consequence 
herbertsmithite has lost some of its initial promise as an almost perfect, spin-1/2,
isotropic kagome HAF.

A more recently discovered candidate for that role is another member of
the atacamite family, namely the polymorph kapellasite, 
$\alpha$-Cu$_3$Zn(OH)$_{6}$Cl$_{2}$,\cite{Colman:2008,Janson:2008_Kagome_Kapellasite,Colman:2010,Fak:2012}
of herbertsmithite. Although they share the same
chemical composition, the two minerals have a different crystallographic structure.
Interestingly, however, they both display distinct kagome structures, although
in different ways.  Thus, in kapellasite the spin-1/2 kagome lattice is
obtained by the regular doping of a 2D triangular Cu$^{2+}$ metal-site sublattice 
with diamagnetic Zn$^{2+}$ ions.  By contrast, in herbertsmithite it is
obtained by a similar diamagnetic dilution with Zn$^{2+}$ ions of the
3D pyrochlore-like sublattice.  It is asserted\cite{Fak:2012} for
kapellasite that while the Cu/Zn mixing leads to some intralayer disorder within the 
kagome planes, it cannot induce any appreciable interlayer coupling,
unlike in herbertsmithite.  

It should be noted, however, that a theoretical electronic study using
density functional theory (DFT) within the local density
approximation,\cite{Janson:2008_Kagome_Kapellasite} of both the material kapellasite and its relative
haydeeite,
Cu$_3$Mg(OH)$_{6}$Cl$_{2}$, has
revealed significant non-NN exchange coupling strengths, especially
those corresponding to bonds across the diagonals of the hexagons on
the kagome lattice.  Furthermore, recent high-temperature series
expansion fits to the measured DC magnetic susceptibility, $\chi_{\rm
  DC}(T)$, as a function of temperature $T$, for
kapellasite,\cite{Fak:2012} seem to give a nearest-neighbor (NN)
exchange interaction ($J_1$) on the kagome planes that is 
{\it ferromagnetic} in nature (i.e., $J_{1}<0$), with the overall
antiferromagnetic behavior of the material explained by large positive
further-neighbor interactions.

A spatially anisotropic version of the spin-1/2 kagome HAF has also been suggested 
to have been realized experimentally in the minerals vorborthite, 
Cu$_{3}$V$_{2}$O$_{7}$(OH)$_{2}$$\cdot$2H$_{2}$O,\cite{Hiroi:2001_kagome_Volborthite,
Bert:2005,Yoshida:2009,Yamashita:2010_kagome_Volborthite,Okamoto:2011,Yoshida:2011,Wulferding:2012_kagome} and 
vesignieite, BaCu$_{3}$(VO$_{4}$)$_{2}$(OH)$_{2}$.\cite{Okamoto:2011,Wulferding:2012_kagome,Okamoto:2009,
Colman:2011,Quilliam:2011}  In both of these materials the Cu sites form a slightly
distorted kagome network, resulting in two inequivalent Cu sites per triangle,
namely one Cu1 site and two Cu2 sites per triangle.  
For example, volborthite has a monoclinic distortion that 
deforms the equilateral triangles of the isotropic kagome network into isosceles triangles.
In this material the difference in the Cu1--Cu2 and Cu2-Cu2 bond lengths is about 3\%.
In turn, theoretical modelling of the thermodynamic properties then leads to a suggested
difference between two of the NN (Cu1--Cu2) exchange
constants ($J_1$) and the third (Cu2--Cu2) one ($J_{1}'$) on each triangle of the kagome 
lattice.  In volborthite this magnetic anisotropy is around 20\%.  
The anisotropy is much less pronounced in vesignieite where the difference 
in bond length is less than 0.1\%, and the material is closer to being structurally
isotropic.  Another recently discovered spin-1/2 deformed kagome-lattice antiferromagnet is 
the material Rb$_{2}$Cu$_{3}$SnF$_{12}$.\cite{Morita:2008,Ono:2009,Matan:2010}

Although both volborthite and vesignieite have a reduced symmetry compared
with the structurally perfect herbertsmithite, they do offer some advantages.  As noted above,
this latter compound shows antisite disorder with up to 10\% of the magnetic 
Cu$^{2+}$ ions exchanged by Zn$^{2+}$ ions, thereby leading to a weak interlayer
magnetic coupling between the kagome planes as well as magnetic vacancies within them.
By contrast, in both volborthite and vesignieite their intermediate layers
between the kagome planes contain V$^{5+}$ ions, and hence antisite disorder of the
Cu ions is prevented.  Although vesignieite is much less anisotropic 
than volborthite it suffers in practice, like herbertsmithite, from low sample 
quality.  One of the main experimental advantages of studying
the more anisotropic volborthite over either of herbertsmithite or vesignieite is
that it is much easier to prepare with fewer impurities.  Nevertheless, we note 
that the nature of the magnetic couplings in this material has been
questioned in a recent study,\cite{Janson:2010} where it is pointed out
that the local environments of the two inequivalent types of Cu sites differ
essentially in important ways.  DFT is then used to show
that volborthite should not be modeled as an anisotropic $J_{1}$--$J_{1}'$ kagome-lattice HAF,
but rather as a $J_{1}'$--$J_{2}'$--$J_{1}$ model, more reminiscent of coupled {\it frustrated} chains, in 
which two-thirds of the kagome sites (viz., the Cu2 sites) are considered as $J_{1}'$--$J_{2}'$ chains 
(i.e., with {\it ferromagnetic} NN exchange, $J_{1}' < 0$ and frustration induced by 
next-nearest-neighbor (NNN) exchange, $J_{2}' > 0$), and with the chains coupled via NN 
exchange bonds of strength $J_1$ between the Cu2 and remaining Cu1 sites.

\begin{figure*}
\subfigure[]{
\includegraphics[width=4cm]{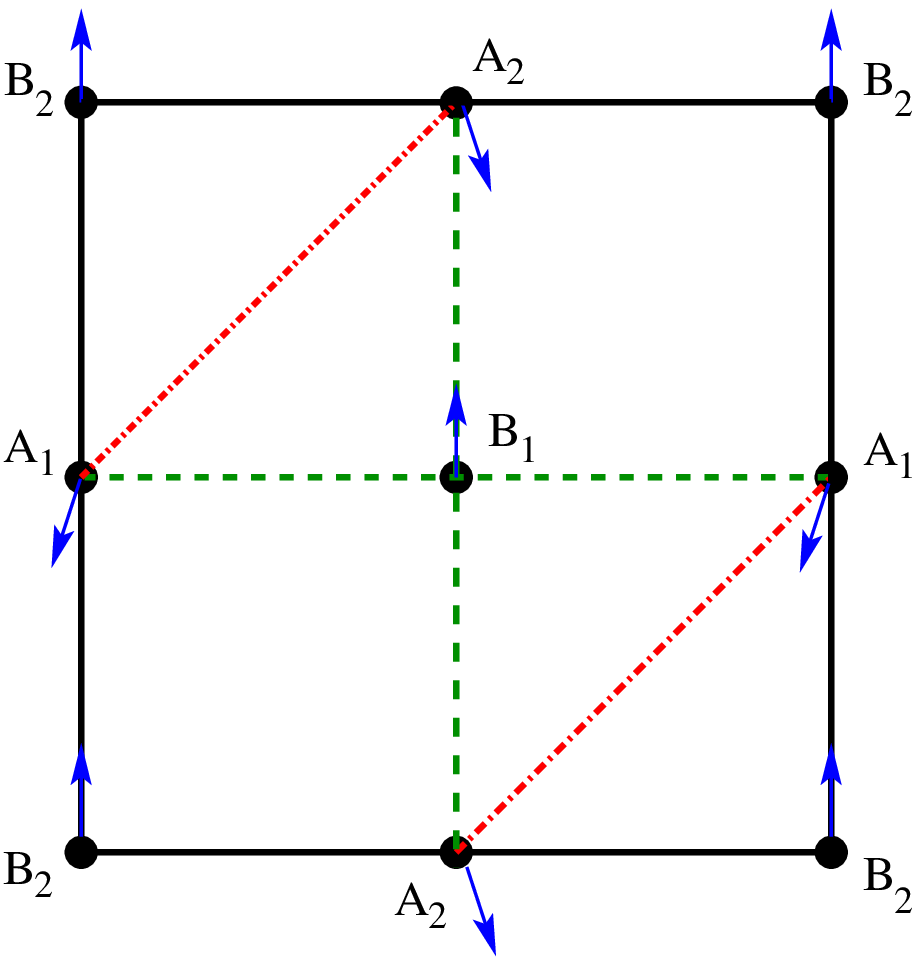}
}%
\hspace{0.3cm}
\subfigure[]{\includegraphics[width=5.5cm]{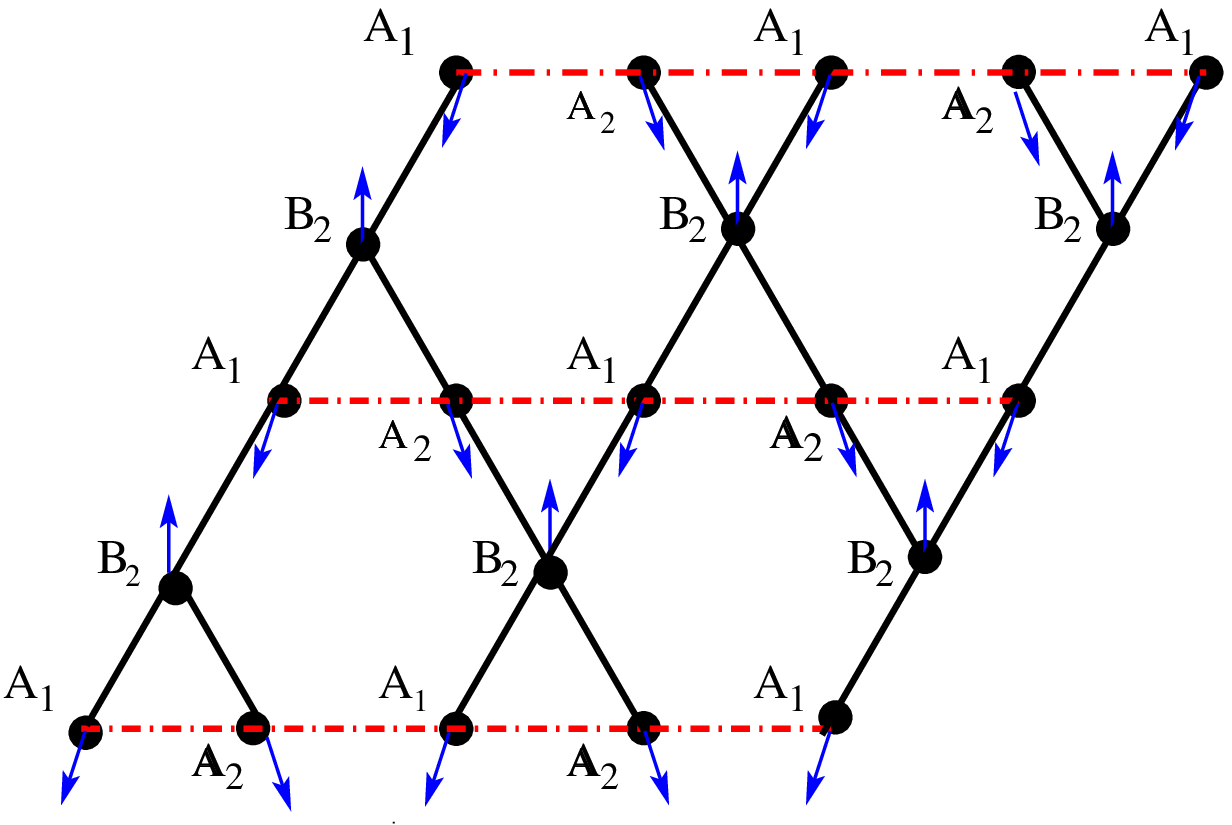}
}%
\hspace{-1cm}
\subfigure[]{\includegraphics[width=5.5cm]{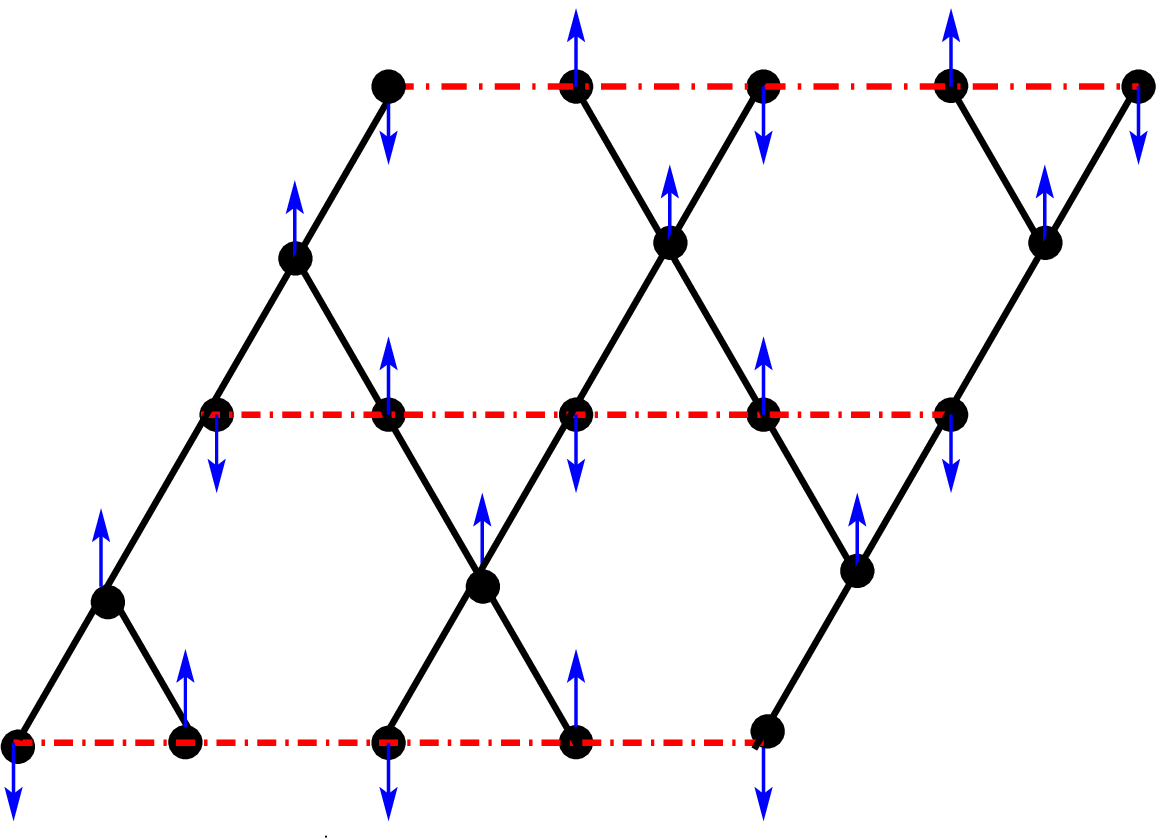}
}%
\caption{
  (Color online) (a) The interpolating kagome-square model with (black) solid bonds --- 
  $J_{1}$, (green) dashed bonds - - $J_{1}'$, and (red) dash-dot bonds - $\cdot$ - $J_{2}$, showing the canted state; 
  and the equivalent anisotropic kagome model when $J_{1}'=0$, showing (b) the coplanar ferrimagnetic canted state 
  and (c) the collinear ferrimagnetic semi-striped state. In all cases the (blue) arrows represent spins
  located on lattice sites $\bullet$.
  }
\label{fig1}
\end{figure*}

Although it is thus often very uncertain as to whether a given real material does or
does not provide an experimental realization of a particular theoretical model, such
as the kagome-lattice HAFs considered here, there is still
much to be gained by a systematic theoretical comparison between such models.
It is of particular interest in such comparisons to use, wherever possible,
the same theoretical technique.  Among the relatively few widely applicable and
systematically improvable (within a well-defined hierarchical approximation scheme)
such tools is the coupled cluster method (CCM).\cite{Bi:1991,Bishop:1998,Ze:1998,
Farnell:2002,Fa:2004}  Our intention here is to use the CCM to further the 
study of the HAF, with NN interactions only, on an anisotropic kagome 
lattice.\cite{Apel:2007,Yavo:2007,Wang:2007,Kagome_Schn:2008,Nakano:2011b,Zyuzin:2012}

By now the CCM has been used to study a huge number of quantum spin-lattice
problems (see, e.g., Refs.~\onlinecite{Fa:2001,Bishop:2010_KagomeSq,
Gotze:2011,Ze:1998,Farnell:2002,Fa:2004,Bi:1993,Zeng:1996,Bi:1998_PRB58,Kr:2000,
Bishop:2000,Fa:2001_PRB64,rachid05,Schm:2006,Fa:2008,Bi:2008_PRB,darradi08,Bi:2008_JPCM,
Bi:2008_EPL,Bi:2009_square_triangle,Farnell:2009_VB-CCM,Richter:2010,
Bishop:2010_UJack,Reuther:2011_J1J2J3mod,DJJF:2011_honeycomb,PHYLi:2012_honeycomb_J1neg,
Bishop:2012_honeyJ1-J2,Bishop:2012_checkerboard} and references cited therein). 
Among these, of particular interest here are applications of the 
CCM to the frustrated spin-1/2 $J_{1}$--$J_{2}$ HAF model on the square 
lattice\cite{Bi:1998_PRB58,Bi:2008_PRB,darradi08,Bi:2008_JPCM,Richter:2010}
with NN bonds (of strength $J_{1} > 0$) competing
with NNN bonds (of strength $J_{2} > 0$), and various models related to it
by removal of some of the NNN $J_2$ bonds.  When half of the $J_2$ bonds are removed 
these include an interpolating square-triangle HAF (or spatially anisotropic 
triangular HAF),\cite{Bi:2009_square_triangle} the Union Jack lattice
model,\cite{Bishop:2010_UJack} and the anisotropic planar pyrochlore (or
checkerboard) HAF (also known as the crossed chain model).\cite{Bishop:2012_checkerboard}

A further modification of the original square-lattice $J_{1}$--$J_{2}$
model is now to remove another half of the $J_{2}$ bonds in such a way as to leave
half of the fundamental square plaquettes with one $J_{2}$ bond and the other
half with none. One way of doing this in a regular fashion results in
the Shastry-Sutherland model\cite{Shastry:1981} in which no $J_{2}$ bonds
meet at any lattice site and every site is five-connected (by four NN
$J_{1}$ bonds and one $J_{2}$ bond).  The CCM has also been successfully applied to this
model.\cite{rachid05,Farnell:2009_VB-CCM}  

Another similar such model, of particular interest here,
arises from removing alternate diagonal lines of $J_2$ bonds from the interpolating
square-triangle HAF (which itself arises from the square-lattice $J_{1}$--$J_{2}$ HAF
model by removing all of the diagonal lines of $J_2$ bonds in the same direction).  It differs
from the Shastry-Sutherland model primarily in that the square lattice now breaks into 
two square-sublattices of A sites and B sites, respectively, such that the A sites are all 
six-connected (by four NN $J_1$ bonds and two NNN $J_2$ bonds), while the B sites are
all four-connected (by four NN $J_1$ bonds only).  A particularly relevant generalization
of the model for present purposes arises from introducing an additional anisotropy
in the NN bonds such that along alternating rows and columns the NN bonds are allowed to
have the strength $J_{1}'$, as shown in Fig.~\ref{fig1}(a).
Clearly, when $J_{1}'=0$ the model reduces to the anisotropic kagome-lattice HAF,
shown equivalently in Fig.~\ref{fig1}(b), which is the subject of the present paper.

The spin-1/2 interpolating kagome-square model described above and shown
in Fig.~\ref{fig1}(a) was studied by us in an earlier paper, using the 
CCM.\cite{Bishop:2010_KagomeSq} In that paper we were mainly interested in the
quantum phase transition line in the $J_{1}'$--$J_2$ plane (with $J_{1} \equiv 1$) 
in the model between the two quasiclassical states with
antiferromagnetic N\'{e}el order and ferrimagnetic canted order.  By
contrast, in the present paper we focus purely on the $J_{1}' \equiv 0$ case
corresponding to the anisotropic kagome-lattice HAF that has been suggested
as one possibility to describe the magnetic properties of volborthite and vesignieite,
as discussed above.  Our main aim is not so much to shed light on the structure of
the paramagnetic GS phase of the isotropic spin-1/2 kagome-lattice HAF, but to
determine the boundaries of this phase as the anisotropy is varied.

After first describing the model in
Sec.~\ref{model_section}, we apply the CCM to investigate its
GS properties. The CCM is itself described briefly in
Sec.~\ref{CCM}, and our results are then presented in
Sec.~\ref{results}.  We conclude in Sec.~\ref{discussions_conclusions}
with a discussion of the results.

\section{The Model}
\label{model_section}
As mentioned in Sec.~\ref{intro}, in a previous paper\cite{Bishop:2010_KagomeSq}
we considered a depleted (and anisotropic) variant of the archetypal and much-studied
$J_{1}$--$J_{2}$ model in which three-quarters of the $J_{2}$ bonds
are removed from it in the pattern shown in Fig.~\ref{fig1}(a).
The square-lattice representation of the model shown in
Fig.~\ref{fig1}(a) contains the two square sublattices of A sites and
B sites respectively, and each of these in turn contains the two
square sublattices of ${\rm A}_{1}$ and ${\rm A}_{2}$ sites, and 
${\rm B}_{1}$ and ${\rm B}_{2}$ sites respectively, as shown. It is very
illuminating to consider the anisotropic variant (viz., the interpolating 
kagome-square model or $J_{1}$--$J_{1}'$--$J_{2}$ model) in which half of the
$J_{1}$ bonds are allowed to have the strengths $J_{1}' > 0$ along
alternating rows and columns, as shown in Fig.~\ref{fig1}(a). 
All of the bonds joining sites $i$ and
$j$ are of standard Heisenberg type, i.e., proportional to 
{\bf s}$_{i}$$\cdot${\bf s}$_{j}$, where the operators 
{\bf s}$_{k}=(s_{k}^{x},s_{k}^{y},s_{k}^{z})$ are the quantum spin
operators on lattice site $k$, with {\bf s}$_{k}^{2}=s(s+1)$ and
$s=\frac{1}{2}$ for the quantum case considered here. 

The spin-1/2 HAF's on the 2D kagome and square lattices are
represented respectively by the limiting cases \{$J_{1}=J_{2},
J_{1}'=0$\} and \{$J_{1}=J_{1}', J_{2}=0$\}. The limiting case
\{$J_{1}=J_{1}'=0; J_{2}>0$\} represents a set of uncoupled one-dimensional (1D) HAF
chains. The case $J_{1}'=0$ with $J_{2} \neq J_{1}$ represents a
spatially anisotropic kagome HAF considered
recently by other authors,\cite{Apel:2007,Yavo:2007,Wang:2007,Kagome_Schn:2008,
Nakano:2011b,Zyuzin:2012} especially in the quasi-1D
limit where $J_{2}/J_{1} \gg 1$.\cite{Kagome_Schn:2008,Zyuzin:2012} It is this latter
model where $J_{1}'=0$ that is considered here (although, for technical reasons,
we note that we actually set $J_{1}'$ to be a small positive value, henceforth chosen
to be $J_{1}'=10^{-5}$).  Our model is thus equivalently shown in the kagome-lattice
geometry of Fig.~\ref{fig1}(b).  It thus comprises parallel chains of spins
on A-sites coupled in a NN fashion along the chains by bonds of strength $J_2$, 
with the chains then cross-linked
via B$_2$ sites with NN bonds of strength $J_1$.  The totality of sites
arranged in a regular kagome lattice on which each component triangle is
comprised of two $J_1$ bonds and one $J_2$ bond.  

The Hamiltonian of the resulting
anisotropic kagome-lattice model is thus
\begin{equation}
H = J_{1}\sum_{\substack{\langle i,j \rangle \\ i \in {\rm A}, j \in {\rm B}_{2}}} {\bf s}_{i}\cdot {\bf s}_{j} + 
    J_{2}\sum_{\substack{\langle i,j \rangle \\ i \in {\rm A_{1}}, j \in {\rm A}_{2}}} {\bf s}_{i}\cdot {\bf s}_{j}\,, 
\label{hamiltonian}
\end{equation}
where the sum on $\langle i,j \rangle$ runs over all NN pairs (of the sort specified in 
each sum), counting each bond once and once only.  Henceforth we consider the 
model where the spins on all lattice sites have spin quantum number $s=\frac{1}{2}$, and
where both types of bonds are antiferromagnetic in nature (i.e., $J_{1} \geq 0, J_{2} \geq 0$)
and hence act to frustrate one another.  With no loss of generality we may then choose the energy scale
by setting $J_{1} \equiv 1$.  We are interested in the infinite-lattice limit, 
$N_{K} \equiv \frac{3}{4}N \to \infty$, where $N_K$ is the number of sites on
the kagome lattice and $N$ is the number of sites on the square lattice of
Fig.~\ref{fig1}(a) before the ($\frac{1}{4}N$) B$_1$ sites have been removed.

Considered as a classical model (corresponding to the case where the
spin quantum number $s \to \infty$) the interpolating
kagome-square model of Fig.~\ref{fig1}(a) (i.e., with $J_{1}' \neq 0$) 
has only two GS phases separated by
a continuous (second-order) phase transition at 
$J_{2}=J_{2}^{{\rm cl}} \equiv \frac{1}{2}(J_{1}+J_{1}')$. For 
$J_{2} < J_{2}^{{\rm cl}}$ the system is N\'{e}el-ordered on the square lattice, while
for $J_{2} > J_{2}^{{\rm cl}}$ the system has noncollinear (but coplanar) canted
order as shown in Fig.~\ref{fig1}(a), in which the spins on each of
the A$_{1}$ and the A$_{2}$ sites are canted respectively at angles
($\pi \mp \phi$) with respect to those on the B sublattice, all of the
latter of which point in the same direction. The lowest-energy state
in the canted phase is obtained with $\phi = \phi_{{\rm cl}} \equiv
\cos^{-1}(J_{2}^{{\rm cl}}/J_{2}$). The N\'{e}el state, for
$J_{2} < J_{2}^{{\rm cl}}$, simply corresponds to the case 
$\phi_{{\rm cl}} = 0$. 

For the case of the anisotropic kagome-lattice HAF considered here 
(i.e., with $J_{1}' = 0$), the classical 
($s \to \infty$) ground states are those spin configurations that satisfy the 
condition that for each elementary triangular plaquette of the kagome lattice in 
Fig.~\ref{fig1}(b) the energy is minimized.  If we take the angle $\phi$ to
be such that the middle spin (viz., that on a B$_2$ site) of a given triangular 
plaquette forms angles ($\pi \pm \phi$) with the other two (chain) spins of the same
plaquette (viz., those on A$_2$ and A$_1$ sites respectively), the total 
energy of the lattice, for classical spins of length $s$, is
$E = \frac{2}{3}N_{K}s^{2}[2J_{1} \cos (\pi-\phi) + J_{2} \cos (2\phi)]$. 
 
For $J_{2} < \frac{1}{2}J_{1}$ this energy is minimized with $\phi=0$, and
the classical GS phase is thus collinear and unique, with
the spins (on the A sites) along the $J_{2}$-bond chains aligned in one direction and
the remaining spins (on the B$_2$ sites) on the kagome lattice aligned in the opposite
direction.  As a convenient shorthand notation we henceforth refer to this collinear 
ferrimagnetic state as the N\'{e}el$'$ state.  Indeed, the N\'{e}el$'$ state of 
the anisotropic kagome-lattice HAF is precisely equivalent to the N\'{e}el state of the 
interpolating kagome-square model of Fig.~\ref{fig1}(a) (i.e., before the removal of 
the B$_1$ sites in the limiting case $J_{1}'=0$) from which it is derived. 
The total spin of this classical collinear N\'{e}el$'$ ferrimagnetic state is thus 
$S_{{\rm tot}}=\frac{1}{3}N_{K}s$ where each spin has magnitude $s$. 
In terms of the saturation magnetization (i.e., in the ferromagnetic 
state with all spins aligned in the same direction), $M_{{\rm sat}} \equiv N_{K}s$, 
the total magnetization in this collinear ferrimagnetic
state is $M^{\rm tot}= \frac{1}{3}M_{\rm sat}$.  For the quantum case the 
Marshall-Lieb-Mattis theorem\cite{Marshall:1955,Lieb:1962} may also be
used to show, for the limiting case $J_{2}=0$ only, that the exact ground
state has the same value $S_{{\rm tot}}=\frac{1}{3}N_{K}s$ of the
total spin as its classical counterpart.  

By contrast, for $J_{2} > \frac{1}{2}J_{1}$, the classical GS energy is minimized 
with the canting angle $\phi = \phi_{{\rm cl}} \equiv \cos^{-1}(\frac{J_1}{2J_{2}})$.
We expect that coplanar canted states will then be favored by either thermal or quantum
fluctuations, and henceforth we only consider coplanar states from among this degenerate
manifold that includes noncoplanar states.  The total magnetization of this canted 
ferrimagnetic state is $M^{\rm tot}= \frac{1}{3}(2\cos \phi - 1)N_{K}s = \frac{1}{3}(\frac{J_1}{J_2}- 1)M_{\rm sat}$.  
The ground state of the HAF on the isotropic kagome
lattice (i.e., with $J_{2}=J_{1}$) falls in this regime, and has the
canting angle $\phi=\frac{\pi}{3}$ demanded by symmetry.  Only for this case
does the total classical magnetization vanish, $M^{\rm tot}= 0$.  The
classical ensemble of degenerate coplanar states is now characterized
by two variables for each triangular plaquette, namely the angle
$\phi$, and the two-valued chirality variable $\chi=\pm 1$ that
defines the direction (anticlockwise or clockwise) in which the spins
turn as one transverses the plaquette in the positive (anticlockwise)
direction. For a given value of $J_{2} > \frac{1}{2} J_{1}$ 
the different degenerate canted states arise from the various
possible ways to assign positive or negative chiralities to the
triangular plaquettes of the lattice.  (Appendix A of Ref.~\onlinecite{Yavo:2007} 
gives a good description of the constraints that these chiralities
need to satisfy.)  

We note that the well-known $q=0$ and $\sqrt{3} \times \sqrt{3}$
states of the (isotropic) kagome-lattice HAF are the special cases, respectively,
where all of the chiralities are the same, and where basic triangular plaquettes
joined by a vertex have opposite chiralities.  The $q=0$ state is one in which
the spins on each of the three sublattices (of A$_1$, A$_2$, and B$_2$ sites
respectively) are parallel to one another, and make an angle of 
120$^\circ$ with the spins on the other two sublattices, while the 
$\sqrt{3} \times \sqrt{3}$ state contains nine sublattices and is obtained by deleting 
$\frac{1}{4}$ of the sites (viz., the B$_1$ sites) of the ordered spins of a triangular lattice 
to form the kagome lattice (and see also, e.g., Refs.~\onlinecite{Harris:1992,Chubukov:92,
Tay:2011a} for further details).  Clearly, the limiting case $J_{1}' \to 0$ of the classical
ground state of the interpolating kagome-square model shown in Fig.~\ref{fig1}(a) is just
the state shown in Fig.~\ref{fig1}(b), which corresponds to the $q=0$ state for the
isotropic ($J_{2}=J_{1}$) kagome-lattice HAF.

The HAF on the isotropic kagome lattice (i.e., with 
$J_{2}=J_{1}$) is especially interesting since for this case, with
$\phi=\frac{\pi}{3}$, the number $\Omega$ of degenerate classical spin
configurations grows exponentially with the number $N_{K}$ of spins,
so that even at zero temperature the system has a nonzero value of the
entropy per spin.  A previous high-order CCM study\cite{Gotze:2011} of the isotropic 
kagome-lattice HAF showed that for the extreme quantum case, $s=\frac{1}{2}$,
the $q=0$ state is energetically favored over the $\sqrt{3} \times \sqrt{3}$ 
state, while for any $s>\frac{1}{2}$ the $\sqrt{3} \times \sqrt{3}$ state 
is selected over the $q=0$ state.  For both the $\sqrt{3} \times \sqrt{3}$ 
and the $q=0$ states it was further found that the magnetic order is
strongly suppressed by quantum fluctuations.  In particular, the order
parameter (viz., the average local on-site magnetization or sublattice 
magnetization) $M_K$ was found to vanish
for both $s=\frac{1}{2}$ and $s=1$, while nonzero values for $M_K$ were found
for $s=\frac{3}{2}$, 2, $\frac{5}{2}$, and 3. 

By contrast, for the anisotropic case (with  $J_{2} \neq J_{1}$), 
the classical degeneracy $\Omega$ has
been shown\cite{Yavo:2007} to grow exponentially with $\sqrt{N_{K}}$ 
[i.e., $\Omega \varpropto {\rm exp} (c\sqrt{N_{K}})]$, so that the GS
entropy per spin vanishes in the thermodynamic limit. Clearly, since
in the limit $J_{2} \to J_{1}$ the anisotropic model approaches
the isotropic model, the anisotropic model must have an appropriately
large number of low-lying excited states that become degenerate with
the ground state in the isotropic limit, $J_{2} \to J_{1}$.

Of course, it is not clear, once the kagome lattice is allowed to
become anisotropic (1.e., with $J_{2} \neq J_{1}$), 
that the $q=0$ state should necessarily remain the lowest-energy
state among the (sub-extensive) ensemble of classically degenerate states.
Continuity would clearly suggest, however, that this should be the case as long
as $J_2$ is not too different from $J_1$.  We have partially checked this within
our own CCM calculations by showing that the the $q=0$ state remains lower in energy
than the $\sqrt{3} \times \sqrt{3}$ state, for example, over the 
range of anisotropicity studied here.

We also note that as $J_{2} \to \infty$ the classical canting angle 
$\phi_{{\rm cl}} \to \frac{1}{2}\pi$, and the spins on the A sublattice chains become
antiferromagnetically ordered, as is expected, and these spins are
orientated at 90$^{\circ}$ to those on the B$_2$
sublattice.  For the particular ordering in Fig.~\ref{fig1}(b), which arises
from that in Fig.~\ref{fig1}(a) in the limit $J_{1}' \to 0$, the spins
on the B$_2$ sublattice are themselves parallel and hence ferromagnetically
ordered.  Of course there is complete degeneracy at the classical
level in this decoupled-chain limit (i.e., when $J_{2} \to \infty$) 
between all states for which the relative ordering
directions for spins on the A and B$_2$ sublattices are arbitrary. In the same 
limit the quantum spin-1/2 problem considered here should also comprise decoupled
antiferromagnetic chains on the A-sublattice sites. We expect that this
degeneracy in relative orientation might be lifted by quantum
fluctuations by the well-known phenomenon of {\it order-by-disorder}.\cite{Vi:1977} 
Since it is also true that quantum
fluctuations generally favor collinear ordering, a preferred state is
thus likely to be the so-called ferrimagnetic semi-striped state shown
in Fig.~\ref{fig1}(c) where the A sublattice is now N\'{e}el-ordered
in the same direction as the B$_2$ sublattice is ferromagnetically
ordered.  (Note that in the square-lattice geometry of Fig.~\ref{fig1}(a)
and where the B$_1$ sites and the $J_{1}'$ bonds are retained,
alternate rows (and columns) are thus ferromagnetically and
antiferromagnetically ordered in the same direction in the
semi-striped state, which is the origin of its name.)

\section{Coupled Cluster Method}
\label{CCM}
We now apply the CCM (see, e.g., Refs.~\onlinecite{Bi:1991,Bishop:1998,Ze:1998,
Farnell:2002,Fa:2004} and references cited therein) to the spin-1/2 
anisotropic kagome-lattice HAF discussed in Sec.~\ref{model_section} above.  
At a very general level the method provides one of the most versatile techniques 
now available in quantum many-body theory.  At attainable levels of
computational implementation it has been shown to provide some of the
most accurate results ever obtained for a large number of quantum many-body 
systems in quantum chemistry, as well as in condensed matter, atomic, molecular, 
nuclear, and subnuclear physics.\cite{Bi:1991,Bishop:1998}  More specifically 
for present purposes, it has been very successfully applied by now
to a large number of systems of interest in quantum magnetism (see, e.g., 
Refs.~\onlinecite{Fa:2001,Bishop:2010_KagomeSq,
Gotze:2011,Ze:1998,Farnell:2002,Fa:2004,Bi:1993,Zeng:1996,Bi:1998_PRB58,Kr:2000,
Bishop:2000,Fa:2001_PRB64,rachid05,Schm:2006,Fa:2008,Bi:2008_PRB,darradi08,Bi:2008_JPCM,
Bi:2008_EPL,Bi:2009_square_triangle,Farnell:2009_VB-CCM,Richter:2010,
Bishop:2010_UJack,Reuther:2011_J1J2J3mod,DJJF:2011_honeycomb,PHYLi:2012_honeycomb_J1neg,
Bishop:2012_honeyJ1-J2,Bishop:2012_checkerboard} and references cited therein), as 
we have already noted in Sec.~\ref{intro}.

The method of applying the CCM to quantum magnets has been described
in detail elsewhere (see, e.g.,
Refs.~[\onlinecite{Ze:1998,Kr:2000,Fa:2001,Schm:2006,Fa:2004}]
and references cited therein).  It relies on building multispin
correlations on top of a suitably chosen, normalized, GS model (or reference) state $|\Phi\rangle$ in a
systematic hierarchy of approximations that we described below.  The reference state
$|\Phi\rangle$ is required only to be a fiducial vector for the system in the sense
that all possible states of the system can be described in terms of it as a linear
combination of states obtained from it by acting on it with members of some suitably chosen complete
set of mutually commuting multispin creation operators, $C_I^+ \equiv (C^{-}_{I})^{\dagger}$.  In
this way $|\Phi\rangle$ acts as a generalized vacuum state with respect to the
set of operators $\{C_I^+\}$.  It is often chosen as a classical ground state of the model
under investigation, and for the present anisotropic kagome-lattice HAF we use mainly
the canted (coplanar) ferrimagnetic state shown in Fig.~\ref{fig1}(b) 
(including the collinear N\'{e}el$'$ state which is its limiting form when
the canting angle $\phi \to 0$), although we also discuss briefly in
Sec.~\ref{discussions_conclusions} the use of the semi-striped
(collinear) ferrimagnetic state shown in Fig.~\ref{fig1}(c) as a CCM model state.

Once the set $\{|\Phi\rangle, C_I^+\}$ has been suitably chosen, the exact GS ket and bra wave 
functions of the system are parametrized within the CCM, in terms of them, in the {\it exponential} forms
that are the hallmark of the method,
\begin{eqnarray}
\label{CCM-ket} 
|\Psi\rangle={\rm e}^S|\Phi\rangle \, , \quad S=\sum_{I\neq 0}{\cal S}_I C_I^+ \, , \\
\label{CCM-bra}
\langle \tilde{\Psi}|=\langle \Phi |\tilde{S}{\rm e}^{-S} \, , \quad \tilde{S}=1+
\sum_{I\neq 0}\tilde{{\cal S}}_IC_I^{-} \, ,
\end{eqnarray}
where we define $C_{0}^{+} \equiv 1$.  It is clear from Eqs.~(\ref{CCM-ket}) and (\ref{CCM-bra}) 
that the normalization has been chosen so that  $\langle \tilde { \Psi}| \Psi \rangle = 
\langle \Phi| \Psi \rangle = \langle \Phi | \Phi \rangle \equiv 1$.  The complete set of
GS CCM correlation coefficients $\{ {\cal S}_{I}, \tilde{{\cal S}}_{I} \}$ ($\forall I\neq 0$) is then
obtained by the requirement that the states $\langle \tilde{\Psi}|$ and  $| \Psi \rangle$ obey
the GS Schr\"odinger equations, $\langle \tilde{\Psi}|H=E\langle \tilde{\Psi}|$
and $H|\Psi\rangle=E|\Psi\rangle$, respectively.  The resulting equations,
\begin{eqnarray}
\label{ket-eq}
\langle\Phi|C_I^-{\rm e}^{-S}H{\rm e}^S|\Phi\rangle = 0 \, , \quad \forall I\neq 0 \, , \\ 
\label{bra-eq}
\langle\Phi|{\tilde S}{\rm e}^{-S}[H, C_I^+]{\rm e}^S|\Phi\rangle = 0 \, , \quad \forall
I\neq 0 \, .
\end{eqnarray}
may, completely equivalently, be derived from the requirement that the GS energy functional
\begin{equation}
\label{energy-functional}
\bar{H} \equiv \langle \tilde{\Psi}|H|\Psi \rangle\, ,
\end{equation}
be stationary with respect to variations in all members of the set 
$\{{\cal S}_{I},\tilde{{\cal S}}_{I}; I\neq 0\}$.  

Once the CCM correlation coefficients have been found from 
solving Eqs.~(\ref{ket-eq}) and (\ref{bra-eq}) it is easy to see that the GS energy 
$E$ is given purely in terms of the ket-state coefficients
$\{{\cal S}_I\}$ as $E=\langle\Phi|{\rm e}^{-S}H{\rm e}^S|\Phi\rangle$.  Clearly, however,
for a more general operator $\hat{O}$, the evaluation of its GS expectation value, 
$\bar{O} \equiv \langle \tilde{\Psi}|\hat{O}|\Psi \rangle$, requires knowledge of 
the set of bra-state correlation coefficients $\{\tilde{{\cal S}}_{I}\}$ as well as 
of the corresponding set of ket-state coefficients $\{{\cal S}_I\}$.

It is very convenient in practice to perform a rotation of the local spin axes of each of 
the spins in the system (i.e., we define a different set of spin axes on every lattice site)
such that all spins in the reference state align along the negative $z$ axes of the local
coordinates.  In practice this simply means that the Hamiltonian has to be re-expressed
in terms of these local axes for each choice of reference state used.  The big advantage
of so doing is that in these local coordinates we have
\begin{equation}
\label{local-spin-frame} 
|{\Phi}\rangle = |\downarrow\downarrow\downarrow\cdots\rangle\, , \quad
C_I^+ = s_{k_1}^{+}s_{k_2}^{+}\cdots s_{k_n}^{+}\, , \; n=1,2,3,\ldots\, ,
\end{equation}
where $s^{+}_{k} \equiv s^{x}_{k} + is^{y}_{k}$, the indices $k_n$ denote arbitrary lattice sites, and the
components of the spin operators are defined in the local rotated coordinate frames. 
We note that for spins of quantum number $s$, each site index $k_n$ in each multispin 
configuration set-index $I=\{k_{1},k_{2},\ldots k_{n}\}$ in
Eq.~({\ref{local-spin-frame}) can be repeated up to a maximum of $2s$ times. Thus, for
the present case, $s=\frac{1}{2}$, all individual spin-site indices in each set-index $I$
are different.  

The magnetic order parameter 
(viz., the average local on-site magnetization) is now given
by $M = -\frac{1}{N} \sum_{i=1}^N \langle\tilde{\Psi}|{s}_i^z|\Psi\rangle$, where
${s}_i^z$ is expressed in the local spin coordinates defined above, 
and $N(\rightarrow \infty)$ is the number of lattice sites.  We denote by $M$ here
the order parameter defined for the general interpolating kagome-square model
of Fig.~\ref{fig1}(a) (i.e., before the $\frac{1}{4}N$ number of B$_1$ sites have 
been removed to yield the anisotropic kagome model).  The corresponding 
order parameter, $M_K$, for the anisotropic kagome-lattice HAF considered here is
similarly given by $M_{K} = -\frac{1}{N_K} \sum_{i=1}^{N_K} \langle\tilde{\Psi}|{s}_i^z|\Psi\rangle$, 
where the sum is taken over only the kagome-lattice sites.  Thus,
in the limiting case $J_{1}'=0$ considered here of the anisotropic kagome-square
model, the spins on the non-kagome B$_1$ sites are frozen to have their spins
aligned exactly along their local negative $z$ axis.  Hence, for the anisotropic
kagome-lattice limit (i.e., when $J_{1}'=0$) of the interpolating kagome-square
model we have the simple relation
\begin{equation}
\label{M-kagome}
M_K = \frac{4}{3}M - \frac{1}{6} \, ,
\end{equation}

The parametrizations of Eqs.~(\ref{CCM-ket}) and (\ref{CCM-bra}) yield,
in principle, the exact GS eigenstate when the complete sets of multispin
creation and destruction operators, $C_I^+$  and $C_I^-$, respectively, is retained.
In practice, of course, it is necessary to make approximations by truncating
the complete set of multispin configuration set-indices $\{I\}$.  In
that case the results for physical quantities such as the GS energy $E$
and order parameter $M$ will naturally depend both on the particular truncation
(i.e., on the configurations specified by the set-indices $I$ that are retained),
as well as on the specific choice of model state to which
those multispin configurations are referred.  We note, however,
that the CCM always exactly obeys the Goldstone linked-cluster theorem 
at every such level of approximation.\cite{Bishop:1998}
Thus, the CCM approach always yields results directly in the thermodynamic limit, 
$N\to\infty$, from the outset, and no finite-size scaling is required.
The results at all levels of approximation are guaranteed to be size-extensive.

Although the CCM is fully (bi-)variational,
as discussed above in connection with the stationarity of $\bar{H}$ in 
Eq.~(\ref{energy-functional}), we note, however, that it does not lead to strict upper bounds for the energy
due to the lack of explicit hermiticity in the parametrizations of the GS ket and bra
wave functions in Eqs.~(\ref{CCM-ket}) and (\ref{CCM-bra}).  This minor drawback is
more than compensated for by the fact that the CCM parametrizations exactly obey the
important Hellmann-Feynman theorem at all levels of approximation.\cite{Bishop:1998}

For the present spin-1/2 model we employ the so-called LSUB$m$
(or lattice-animal-based subsystem) approximation scheme to truncate the expansions 
of $S$ and $\tilde S$ in Eqs.~(\ref{CCM-ket}) and (\ref{CCM-bra}).  In this very widely
tested scheme one includes from the full set of multispin configurations specified
by the set-indices $I$ in Eqs.~(\ref{CCM-ket}), (\ref{CCM-bra}), and   
(\ref{local-spin-frame}) only those involving $m$ or fewer correlated spins in all 
arrangements (or lattice animals in the language of graph theory) which 
span a range of no more than $m$ contiguous lattice  sites.
In this context a set of sites is defined to be contiguous if every site has at 
least one other in the set as a nearest neighbor, and where it is
clearly necessary to include a definition in the geometry (or better, topology) of the 
lattice of which pairs of sites are considered to be NN pairs.  For example,
we choose here to work in the triangular-lattice geometry of the anisotropic
kagome-square model in which the B sublattice
sites of Fig.~\ref{fig1}(a) are defined to have four NN sites joined
to them by either $J_{1}$ bonds or $J_{1}'$ bonds, and the A sublattice
sites are defined to have the six NN sites joined to them by $J_{1}$,
$J_{1}'$, or $J_{2}$ bonds. If we had chosen instead to work in the
square-lattice geometry every site would have four NN sites.  The former
triangular-lattice geometry leads in the limiting case when $J_{1}' = 0$ 
to the natural kagome-lattice geometry of Fig.~\ref{fig1}(b) in which any two of
the A$_1$, A$_2$, and B$_2$ sites forming the basic triangular plaquettes are
considered to be NN pairs.  Clearly, this would not be the case in the latter
square-lattice geometry.  For more details of 
the LSUB$m$ scheme the reader is referred, for example, to Refs.~\onlinecite{Ze:1998,
Farnell:2002,Fa:2004}.

The astute reader will have noted the close connection between the CCM
parametrization of the ket-state wave function given by Eqs.~(\ref{CCM-ket}) and   
(\ref{local-spin-frame}), when the multispin cluster configurations $I$ are restricted 
to those between two spins only (as in the case of the LSUB$m$ scheme used here
with $m=2$, where only NN pair-correlations are included), with that of
(lowest-order) self-consistent spin-wave theory (SWT).\cite{Anderson:1952}
Nevertheless, the two methods are still not identical, due partly to the lack of 
explicit hermiticity between the CCM ket and bra parametrizations, as in
Eqs.~(\ref{CCM-ket}) and (\ref{CCM-bra}), and partly due to the way that 
self-consistency is incorporated within SWT.  For example, in many cases
where SWT is unstable (i.e., gives a negative magnetic order parameter, $M$)
the CCM LSUB$2$ result is usually stable (i.e., gives a positive value for
$M$).  The interested reader is referred to the literature\cite{Xian:2005}
for a detailed discussion of the relationships between SWT and low-order
implementations of the CCM.

Clearly, the LSUB$m$ truncations scheme provides a fully systematic approximation
hierarchy in the sense that each time the truncation index $m$ is increased more
of the Hilbert space is sampled, such that as $m \to \infty$ the approximation 
becomes exact.  Although, as we have already indicated, 
we never need to perform any finite-size scaling, since all CCM 
approximations are automatically performed from the outset in the 
infinite-lattice limit, $N_{K} \to \infty$, where $N_K$ is the number of lattice 
sites, we do still need as a last step in a CCM calculation to extrapolate to 
the exact $m \rightarrow \infty$ limit in the LSUB$m$ truncation index $m$, at 
which the complete (infinite) Hilbert space is reached.  
By now there is a great deal of experience available regarding how one should
extrapolate the GS energy per site $e_{0}(m) \equiv E(m)/N_{K}$ and the magnetic order parameter
$M(m)$.  

Thus, for the GS energy per spin, $e_{0}(m)$, we use the well-tested
empirical scaling ansatz (and see, e.g., Refs.
\onlinecite{Reuther:2011_J1J2J3mod,Kr:2000,rachid05,Schm:2006,
Bi:2008_PRB,Bi:2008_JPCM,darradi08,Bi:2009_square_triangle,Richter:2010,Bishop:2010_UJack,
Bishop:2012_checkerboard}),
\begin{equation}
e_{0}(m)=a_{0}+a_{1}m^{-2}+a_{2}m^{-4}\,.  
\label{Extrapo_E}
\end{equation} 

For the GS magnetic order parameter different extrapolation rules have been used 
depending on the features of the spin-lattice system at hand.  For highly
frustrated spin-lattice systems, a well-tested rule (and see,
e.g., Refs.~\onlinecite{DJJF:2011_honeycomb,PHYLi:2012_honeycomb_J1neg,
Bishop:2012_honeyJ1-J2, Reuther:2011_J1J2J3mod,Bi:2008_PRB,Bi:2008_JPCM,darradi08,Richter:2010,
Bishop:2012_checkerboard}) is
\begin{equation}
M(m) = b_{0}+b_{1}m^{-1/2}+b_{2}m^{-3/2}\,.    
\label{M_extrapo_frustrated}
\end{equation} 
An alternative rule that is useful, for example, for situations when there is some frustration
present, but when it is not too large, is\cite{rachid05,Bishop:2012_checkerboard} 
\begin{equation}
M(m) = c_{0}+c_{1}m^{-\nu}\, ,    
\label{M_extrapo_nu}
\end{equation}
which, as an advantage, leaves the leading exponent open for determination.  The disadvantage
of this scheme is clearly that it involves only the leading two terms in the
asymptotic expansion, compared to the (inherently more accurate) 
three terms used in Eq.~(\ref{M_extrapo_frustrated}).
Clearly, however, when the exponent $\nu$ in Eq.~(\ref{M_extrapo_nu}) is found by fitting to the
CCM LSUB$m$ results to be close to the value 0.5, as is usually found for highly frustrated 
systems, we can then revert to the more accurate form of Eq.~(\ref{M_extrapo_frustrated}).

In the present paper we present
extrapolated results based on LSUB$m$ data sets with $m=\{2,4,6,8\}$.
To check the robustness of the extrapolation rules and to estimate the associated 
error bars in the extrapolations we have also performed extrapolations
using LSUB$m$ data sets with $m=\{2,4,6\}$ and $m=\{4,6,8\}$.  We find in general
that the extrapolated results from all three sets of data are very similar.

The number of independent fundamental LSUB$m$ clusters (i.e., those that are 
inequivalent under the symmetries of the Hamiltonian and of the model state) that 
are retained in the expansions of Eqs.~(\ref{CCM-ket}) and (\ref{CCM-bra}) 
for the CCM correlation operators $S$ and $\tilde{S}$
increases rapidly with the truncation index $m$. For example, the
number of such fundamental clusters for the canted model state of the
interpolating kagome-square model of Fig.~\ref{fig1}(a) is
201481 at the LSUB8 level of approximation in the triangular-lattice
geometry used here where $J_{2}$ bonds are considered to join NN pairs, and this
is the highest level for the present model that we have been able to attain with available
computing power. In order to solve the corresponding coupled sets of
CCM bra- and ket-state equations we use an efficient parallelized CCM code,\cite{cccm} 
and typically employ around 600 processors simultaneously. 

Finally, we note that our CCM calculations based on the canted phase of the
anisotropic kagome-lattice HAF shown in Fig.~\ref{fig1}(b) do not assume for the
$s=\frac{1}{2}$ model considered here that the canting angle $\phi$ takes the same value
$\phi_{{\rm cl}} \equiv \cos^{-1}(\frac{J_1}{2J_{2}})$ as in the classical 
($s \to \infty$) case.  Rather, calculations are first performed 
for an arbitrary choice of canting angle $\phi$.  We then
minimize the corresponding LSUB$m$ approximation for the energy
$E_{{\rm LSUB}m}(\phi)$ with respect to $\phi$ to yield the
corresponding approximation to the quantum canting angle $\phi_{{\rm LSUB}m}$. 
Generally (for $m > 2$) the minimization must be carried
out computationally in an iterative procedure.  Results for the 
canting angle $\phi_{{\rm LSUB}m}$ are presented below in Sec.~\ref{results}.                                                              

\section{Results}
\label{results}
We now present our CCM results for the anisotropic spin-1/2 $J_{1}$--$J_{2}$ HAF 
on the kagome lattice of Eq.~({\ref{hamiltonian}), where we use the canted ferrimagnetic
state shown in Fig.~\ref{fig1}(b) as the CCM model (or reference) state $|\Phi\rangle$
in the representations of Eqs.~(\ref{CCM-ket}) and (\ref{CCM-bra}) of the exact GS ket
and bra wave functions.  Without loss of generality, but simply to set the energy scale,
we henceforth set $J_{1} \equiv 1$.  (Equivalently, when more convenient to do so, we quote
results in terms of the ratio $\kappa \equiv J_{2}/J_{1}$, assuming always
that $J_{1} > 0$.)  We first show in Fig.~\ref{EvsAngle} the GS energy per spin, 
\begin{figure}[t]
\includegraphics[width=6cm,angle=270]{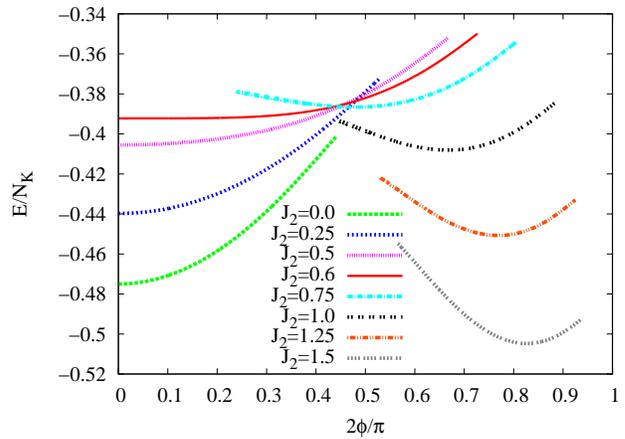}
\caption{
  (Color online) Ground-state energy per spin, $E/N_K$, of the spin-1/2 $J_{1}$--$J_{2}$
  HAF on the anisotropic kagome lattice of Eq.~({\ref{hamiltonian})} (with $J_{1} \equiv 1$), using
  the LSUB4 approximation of the CCM with the ferrimagnetic canted model state shown in 
  Fig.~\ref{fig1}(b), versus the canting angle $\phi$ for various selected values
  of the anisotropy parameter $J_2$.  For $J_{2} \lesssim 0.6$ the minimum is
  at $\phi=0$ (N\'{e}el$'$ order) at this level of approximation, whereas
  for $J_{2} \gtrsim 0.6$ the minimum occurs at $\phi=\phi_{{\rm LSUB}4} \neq 0$, 
  thereby providing evidence of a phase transition at $J_{2} \approx
  0.6$ in this approximation.  We show results for those values of
  $\phi$ for which the corresponding CCM equations having real solutions.
  }
\label{EvsAngle}
\end{figure}
$E/N_{K}$, as a function of the canting angle $\phi$.  Although results are shown
at the LSUB4 level of approximation results for other LSUB$m$ levels are
qualitatively similar.  Curves such as those shown in Fig.~\ref{EvsAngle} show
that at this LSUB4 level of approximation, with $J_{1}=1$,
the GS energy is minimized at $\phi = 0$ for $J_{2} < J_{2}^{{\rm LSUB}4}
\approx 0.60$ and at a value $\phi \neq 0$ for $J_{2} > J_{2}^{{\rm LSUB}4}$.  
Hence, these first results indicate the possibility of a slight shift of the
critical point to $J_{2} = J_{2}^{c_{1}} \equiv J_{2}^{{\rm LSUB}\infty}$ between the quantum ferrimagnetic 
N\'{e}el$'$ and canted phases, from the classical value $J_{2}^{{\rm cl}}=0.5$ when 
$J_{1}=1$.  The fact that N\'{e}el$'$ order survives here, at least at finite
orders of LSUB$m$ approximation, into the regime where it 
would be classically unstable against the formation of canted order is an 
example of a phenomenon that has been observed in many other magnetic systems, namely the 
tendency for quantum fluctuations themselves to favor collinear over noncollinear order.
Nevertheless, we leave till later in this Section a discussion of extrapolating these results
for the critical point $J_{2}^{{\rm LSUB}m}$ to the $m \to \infty$ limit, and hence also
of a comparison of the quantum $s=\frac{1}{2}$ case with its classical 
($s \to \infty$) counterpart.

In Fig.~\ref{angleVSj2} we now show the canting angle $\phi_{{\rm LSUB}m}$ that minimizes the GS
\begin{figure}[t]
\includegraphics[width=6cm,angle=270]{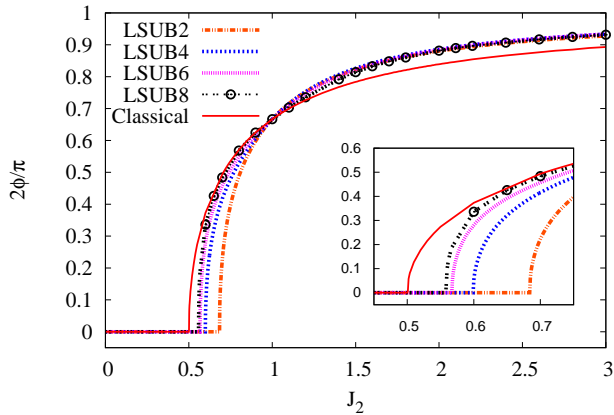}
\caption{
  (Color online) The angle $\phi_{{\rm LSUB}m}$ that minimizes
  the GS energy $E_{{\rm LSUB}m}(\phi)$ of the spin-1/2 $J_{1}$--$J_{2}$
  HAF on the anisotropic kagome lattice of Eq.~(\ref{hamiltonian}) (with $J_{1} \equiv 1$),
  versus the anisotropy parameter $J_2$.  The LSUB$m$ approximations with $m=\{2,4,6,8\}$, using the 
  ferrimagnetic canted state of Fig.~\ref{fig1}(b) as CCM model state, are shown.  The 
  corresponding classical result $\phi_{{\rm cl}} \equiv \cos^{-1}(\frac{1}{2J_{2}})$ 
 is also shown for comparison.
  }
\label{angleVSj2}
\end{figure}
energy $E_{{\rm LSUB}m}$($\phi$) using the ferrimagnetic canted state of 
Fig.~\ref{fig1}(b) as CCM model state, at various CCM
LSUB$m$ levels, with $m=\{2,4,6,8\}$.  
We see from Fig.~\ref{angleVSj2} that, at each LSUB$m$ level shown, the canting
angle $\phi_{{\rm LSUB}m}$ that minimizes the corresponding estimate, 
$E_{{\rm LSUB}m}(\phi)$, for the GS energy $E_{{\rm LSUB}m}(\phi)$ of the model
approaches zero smoothly, but with infinite slope, from the canted state (with 
$\phi_{{\rm LSUB}m} \neq 0$) side of the phase transition, as the anisotropy
parameter $J_2$ is reduced, to the corresponding estimate for the critical point,
$J_{2}^{{\rm LSUB}m}$, and that it then remains zero for all $J_2 < J_{2}^{{\rm LSUB}m}$ 
on the N\'{e}el$'$ state side of the transition.  This behavior is completely
analogous to that seen in the classical version of the model, also shown in
Fig.~\ref{angleVSj2}.  The evidence so far, therefore, is that both the classical and spin-1/2
versions of the HAF on the anisotropic kagome lattice show second-order phase transitions
between ferrimagnetic states with collinear N\'{e}el$'$ order and noncollinear canted
order.  We will see below, however, that this picture changes when the
magnetic order parameter is considered too.

We note first though that, by contrast, the corresponding behavior
observed in the isotropic version (i.e., when $J_{1}'=J_{1} \equiv 1$) of the 
interpolating kagome-square HAF model (or $J_{1}$--$J_{1}'$--$J_{2}$ model) of 
Fig.~\ref{fig1}(a), of which latter model the present HAF model on the anisotropic kagome lattice
is just the special case with $J_{1}'=0$, is quite different.  Thus, in
the spin-1/2 $J_{1}$--$J_{1}'$--$J_{2}$ model with $J_{1}'=J_{1} \equiv 1$, it was 
observed\cite{Bishop:2010_KagomeSq} that at each LSUB$m$ level there is 
a finite jump in $\phi_{{\rm LSUB}m}$ at the corresponding LSUB$m$ approximation
for the phase transition at $J_{2}=J_{2}^{{\rm LSUB}m}$ between the
N\'{e}el state (with $\phi_{{\rm LSUB}m}=0$) and the canted state
(with $\phi_{{\rm LSUB}m} \neq 0$).  This may be compared with the smooth behavior of the classical
canting angle $\phi_{{\rm cl}} \equiv \cos^{-1}(\frac{1}{J_{2}})$ for 
this case for $J_{2} > J_{2}^{{\rm cl}}=1$ for that model. For this latter 
$J_{1}$--$J_{1}'$--$J_{2}$ model in the isotropic case with $J_{1}'=J_{1}$
the evidence was that the phase transition between states with N\'{e}el and canted order
was first-order for the spin-1/2 case compared with its second-order classical
counterpart.  On the other hand one should note that for that case\cite{Bishop:2010_KagomeSq}
we could not completely rule out the possibility that as $m \rightarrow \infty$, with
increasing level of LSUB$m$ approximation, the phase transition at 
$\kappa = \kappa_{c_1} \equiv \kappa_{c_1}^{{\rm LSUB}\infty}$ becomes of second-order type, 
although a weakly first-order one seemed more likely on the basis of the available 
numerical evidence.  

Returning now to the question of estimating the phase transition point at $\kappa = \kappa_{c_1}$
in the present model, we note that previous empirical 
experience\cite{Bishop:2010_UJack,Bishop:2010_KagomeSq} 
shows that the LSUB$m$ estimates at
$\kappa = \kappa_{c_1}^{{\rm LSUB}m}$ fit well to an extrapolation scheme 
$\kappa_{c_1}^{{\rm LSUB}m} = \kappa_{c_1}^{{\rm LSUB}\infty} + cm^{-1}$.  For
the present anisotropic spin-1/2 $J_{1}$--$J_{2}$ model
on the kagome lattice of Eq.~({\ref{hamiltonian}), our phase
transition estimates for $\kappa_{c_1} \equiv \kappa_{c_1}^{{\rm LSUB}\infty}$ are
shown in Table~\ref{table_CritPt}.
\begin{table}[t]
  \caption{
  The critical value $\kappa^{{\rm LSUB}m}_{c_{1}}$ at which the transition 
  between the N\'{e}el$'$ phase ($\phi=0$) and the canted phase ($\phi \neq 0$) occurs in 
  the LSUB$m$ approximation using the CCM with (N\'{e}el or) canted state as model state 
  for the the spin-1/2 $J_{1}$--$J_{2}$ HAF on the anisotropic kagome lattice of 
  Eq.~(\ref{hamiltonian}).
  }
\label{table_CritPt}
\begin{tabular}{ccc} \hline\hline
Method & $\kappa^{{\rm LSUB}m}_{c_{1}}$ &\\ \hline      
LSUB2 & 0.685 &\\
LSUB4 & 0.600 &\\
LSUB6 & 0.568 &\\
LSUB8 & 0.559 &\\  \hline\hline   
\end{tabular}  
\end{table}      
Using the above extrapolation scheme and the whole data set $m=\{2,4,6,8\}$ gives
the estimate $\kappa_{c_1}=0.514 \pm 0.003$, while the corresponding estimate from
using the data set $m=\{4,6,8\}$ is $\kappa_{c_1}=0.515 \pm 0.007$.  In both cases
the quoted error estimates are simply the standard deviations from the associated least-squares fits. 
Our best estimate from combining all of our results is $\kappa_{c_1}=0.515 \pm 0.015$,
which may be compared with the corresponding classical value of
$\kappa_{{\rm cl}} = 0.5$.  Clearly it is not excluded that $\kappa_{c_1}=\kappa_{{\rm cl}}$,
such that the transition point above which collinear N\'{e}el$'$ order disappears occurs at
exactly the same value $\kappa = 0.5$ of the anisotropy parameter for both the
extreme quantum and classical limiting cases of the spin quantum number $s$. 
We also discuss the nature of the phase transition at $\kappa_{c_1}$ for the spin-1/2
model in more detail below.

We note from Fig.~\ref{angleVSj2} that as $J_{2} \rightarrow \infty$ the canting 
angle $\phi \rightarrow \frac{1}{2}\pi$  faster than does the
classical analog $\phi_{{\rm cl}}$.  We also note that for the special case
$J_{2}=1$ (or, equivalently, $\kappa = 1$) of the isotropic kagome lattice, the
CCM LSUB$m$ estimates for the canting angle $\phi$ take the value
$\phi_{{\rm LSUB}m} = \frac{\pi}{3}$ for all values of $m$, as expected by symmetry, and exactly
as in the classical version of the model.

In Fig.~\ref{E}
we show our CCM results for the GS energy per spin, $E/N_{K}$, as a function of $J_{2}$, 
for the present anisotropic spin-1/2 $J_{1}$--$J_{2}$ model
on the kagome lattice of Eq.~({\ref{hamiltonian}).  The CCM LSUB$m$ results with $m=\{2,4,6,8\}$ and
the corresponding extrapolated LSUB$\infty$ results obtained from Eq.~(\ref{Extrapo_E}) are shown.
As explained previously, in each LSUB$m$ approximation we choose the canting angle
$\phi=\phi_{{\rm LSUB}m}$ for each separate value of the parameter $J_2$ that minimizes the 
corresponding CCM estimate for the energy, $E_{{\rm LSUB}m}(\phi)$. 

\begin{figure}[t]
\includegraphics[width=6cm,angle=270]{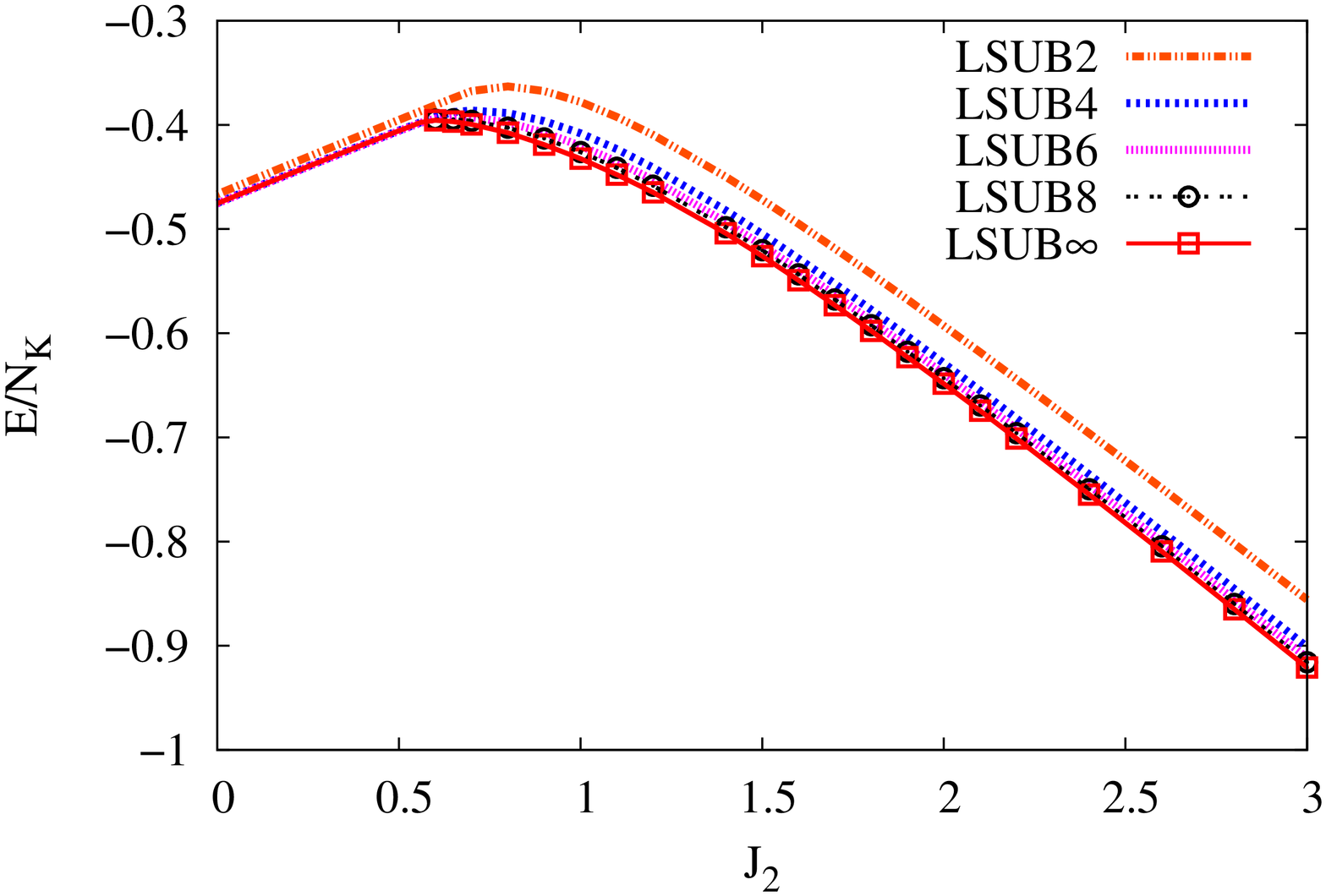}
\caption{
  (Color online) Ground-state energy per spin, $E/N_{K}$, versus $J_{2}$ for the spin-1/2 $J_{1}$--$J_{2}$
  HAF on the anisotropic kagome lattice of Eq.(\ref{hamiltonian}) (with $J_{1} \equiv 1$), using
  the generic ferrimagnetic canted model state shown in Fig.~\ref{fig1}(b) as CCM model state, and
  with the canting angle $\phi=\phi_{{\rm LSUB}m}$ chosen to minimize the corresponding LSUB$m$ estimate
  for the energy, $E_{{\rm LSUB}m}(\phi)$, at each value of $J_2$.  The CCM LSUB$m$
  results with $m=\{2,4,6,8\}$ are shown, together with the corresponding extrapolated LSUB$\infty$
  result from Eq.~(\ref{Extrapo_E}). 
  }
\label{E}
\end{figure}

At the isotropic kagome point (i.e., when $J_{2}=J_{1}=1$) our present best estimate for
the GS energy per spin, based on the extrapolation with the data set $m=\{4,6,8\}$, is
$E/N_{K} \approx -0.4352$.  This is lower than two recent rigorous upper bounds.\cite{Eve:2010,Yan:2011} 
Thus, Evenbly and Vidal\cite{Eve:2010} used the multiscale entanglement renormalization
ansatz to evaluate exactly (up to floating point round-off errors) the energy of a wave function 
of the so-called honeycomb valence-bond crystalline type (with a 36-site unit cell), to give
the rigorous bound $E/N_{K}<-0.4322$.  Similarly, using a simple cluster product state for
the infinite kagome lattice based on a fundamental cluster of 576 sites, for which
the interior of the cluster has the uniform valence-bond patterning expected of a
spin-liquid state, Yan, Huse, and White\cite{Yan:2011} have recently given an improved
rigorous upper bound of $E/N_{K}<-0.4332$.  The same authors\cite{Yan:2011} also use a large-scale
density-matrix renormalization group (DMRG) technique to provide what is certainly one of the most accurate
estimates currently available for the energy per site of the isotropic kagome-lattice HAF, namely
$E/N_{K}=-0.4379 \pm 0.0003$.  This estimate is itself consistent with the best available
large-scale Lanczos exact diagonalization (ED) results for finite clusters of up to $N=42$ 
sites.\cite{Lauchli:2011}  It is also in excellent agreement with that from another very recent 
large-scale DMRG study,\cite{Depenbrock:2012_kagome_spinLiquid} namely $E/N_{K}=-0.4386 \pm 0.0005$.  
Our own present best estimate, cited above, is clearly below
both of the rigorous upper bounds and in good agreement with the DMRG results.  

Our extrapolated result for the energy is also in 
very good agreement with previous CCM estimates\cite{Gotze:2011} for the spin-1/2 HAF on the 
isotropic kagome lattice that used the kagome geometry itself to define the fundamental clusters
of the LSUB$m$ configurations rather than the triangular lattice geometry of Fig.~\ref{fig1}(a) that 
we use here, as discussed above in Sec.~\ref{CCM}, in which the B$_1$ sites are retained. Thus, 
using the kagome-lattice geometry the extrapolated result $E/N_{K} \approx -0.4357$ was found
with the LSUB$m$ set $m=\{4,5,6,7,8,9,10\}$ and $E/N_{K} \approx -0.4372$ with the LSUB$m$ 
set $m=\{6,7,8,9,10\}$.\cite{Gotze:2011}

Figure~\ref{E} shows weak signals of the phase transition point (at 
$\kappa = \kappa_{c_1} \equiv \kappa_{c_1}^{{\rm LSUB}\infty}$) in each of the LSUB$m$ curves,
where a discontinuity in the first derivative of the energy is observed at
the corresponding value $\kappa = \kappa_{c_1}^{{\rm LSUB}m}$.  As usual the transition is 
seen more clearly in the behavior of the average local on-site magnetization, 
$M_{K} \equiv - \frac{1}{N_{K}}\sum^{N_{K}}_{i=1}\langle s^{z}_{i} \rangle$, 
where the sum is taken over all $N_K$ sites of the kagome lattice
and where again the spins are defined in the local, rotated spin axes
in which all spins in the CCM model state point in the negative {\it z}-direction.    

Thus, in Fig.~\ref{M} we show the magnetic order parameter, $M_K$, as a function of $J_{2}$, 
for the present anisotropic spin-1/2 $J_{1}$--$J_{2}$ model on the kagome lattice of 
Eq.~({\ref{hamiltonian}).  
\begin{figure}[t]
\begin{center}
\hspace*{-1.2cm}\includegraphics[angle=270,width=11cm]{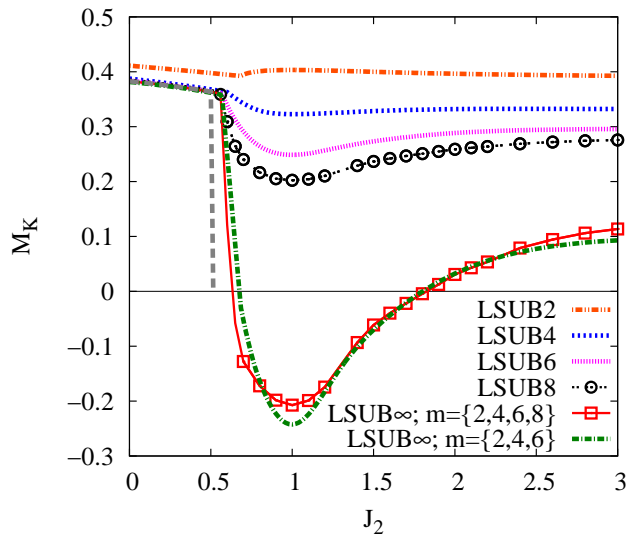}
\caption{
  (Color online) Ground-state magnetic order parameter, $M_K$, versus $J_{2}$ 
  for the spin-1/2 $J_{1}$--$J_{2}$
  HAF on the anisotropic kagome lattice of Eq.~(\ref{hamiltonian}) (with $J_{1} \equiv 1$), using
  the generic ferrimagnetic canted model state shown in Fig.~\ref{fig1}(b) as CCM model state, and
  with the canting angle $\phi=\phi_{{\rm LSUB}m}$ chosen to minimize the corresponding LSUB$m$ estimate
  for the energy, $E_{{\rm LSUB}m}(\phi)$, at each value of $J_2$.  The CCM LSUB$m$
  results with $m=\{2,4,6,8\}$ are shown, together with the corresponding extrapolated LSUB$\infty$
  results using both data sets $m=\{2,4,6,8\}$ and $m=\{2,4,6\}$ for comparison purposes.  
  Extrapolated results are calculated using Eq.~(\ref{M_extrapo_nu}) for the N\'{e}el$'$ state and
  Eq.~(\ref{M_extrapo_frustrated}) for the canted state, respectively; and  
  the (grey) dashed vertical line at $J_{2} = 0.515$ represents our best estimate for the termination point 
  above which collinear N\'{e}el$'$ order disappears, at 
  $J_{2} = J_{2}^{c_{1}} \equiv J_{2}^{{\rm LSUB}\infty}$, as discussed in the text.
  }
\label{M}
\end{center}
\end{figure}
CCM results are shown for LSUB$m$ approximations with $m=\{2,4,6,8\}$, together with various
LSUB$\infty$ extrapolations.  As discussed in Sec.~\ref{CCM}, for the strongly frustrated regime 
in which the canted state is the stable ground state (i.e., for $J_{2} > J_{2}^{c_{1}}$) we use the 
well-tested and established scheme of Eq.~(\ref{M_extrapo_frustrated}), whereas for the
less frustrated regime in which the N\'{e}el$'$ state is the stable ground state (i.e., for
$J_{2} < J_{2}^{c_{1}}$) we use the scheme of Eq.~(\ref{M_extrapo_nu}).  Since for the 
N\'{e}el$'$ state the LSUB$m$ results converge (with increasing values of $m$)
much faster than those for the canted state, as can clearly be seen from Fig.~\ref{M},
it is evident that the use of these different schemes for the two regimes is justified.
Since the approximate transition point at $J_{2} = J_{2}^{{\rm LSUB}m}$ (when $J_{1}=1$) between the two phases
depends slightly on the CCM truncation index $m$, as has already been noted above, and 
as can be seen clearly in Fig.~\ref{M} (and, more explicitly in Table~\ref{table_CritPt}), 
it is clear that the region very near the transition is inherently difficult to extrapolate
accurately.  

To illustrate the sensitivity of our extrapolations to the approximations
used we show in Fig.~\ref{M} the corresponding extrapolations in the two regimes using
both the data sets $m=\{2,4,6,8\}$ and $m=\{2,4,6\}$.  Since the (most accurate) LSUB8
scheme is computationally expensive, results are shown only at the limited set of $J_2$
values indicated by the symbols in Fig.~\ref{M}.  It is very encouraging that the
extrapolated curves for $M_K$ are very steep near $J_{2}^{c_{1}}$ and that they become steeper
still as higher LSUB$m$ approximations are included.  For example, the extrapolated LSUB$\infty$ 
curve for $M_K$ obtained from the set $m=\{2,4,6\}$ becomes zero at $J_{2} \approx 0.67$ (with
$J_{1} \equiv 1$), while that obtained from the set $m=\{2,4,6,8\}$ becomes zero at 
$J_{2} \approx 0.63$ (with $J_{1} \equiv 1$).  It seems reasonable to assume
that the actual LSUB$\infty$ curve will become vertical at the point 
$J_{2} = J_{2}^{c_{1}} \equiv J_{2}^{{\rm LSUB}\infty}$, as is seen in Fig.~\ref{M} 
by the proximity of the extrapolations to the vertical line at $J_{2} = 0.515$, which 
represents our best estimate for the transition point between the N\'{e}el$'$ and canted states.  

Figure~\ref{M} shows the existence of a clear window in the parameter $\kappa$ in which $M_{K} <0$, and hence
in which the canted order present in the model state has vanished.  This (paramagnetic) 
region includes the point $\kappa=1$ corresponding to the isotropic kagome HAF, and it seems
reasonable to assume that the phase present in this regime is the same
as the paramagnetic GS phase of the isotropic model.  As discussed above, the 
lower boundary of this window seems to coincide with the point $\kappa = \kappa_{c_{1}}$ 
above which the collinear N\'{e}el$'$ order disappears.  

Thus, the evidence from the magnetization
data shown in Fig.~\ref{M} now indicates that the transition at $\kappa = \kappa_{c_{1}}$ is
actually between the N\'{e}el$'$ and paramagnetic phases, rather than between the N\'{e}el$'$ and 
canted phases as in the classical case at the corresponding value $\kappa =\kappa_{{\rm cl}}$. 
Nevertheless, we cannot completely exclude the possibility of a very narrow
strip of canted phase between the N\'{e}el$'$ and paramagnetic phases confined to the region
$0.5 < \kappa \lesssim 0.6$.  

We denote by $\kappa_{c_{2}}$ the corresponding critical value
of $\kappa$ that marks the upper boundary of the window in which $M_{K} <0$, and that hence
marks the transition between the paramagnetic and canted phases.  From Fig.~\ref{M} we find
estimates $\kappa_{c_{2}} \approx 1.83 \pm 0.02$ from the extrapolated LSUB$\infty$ curve using
the data set $m=\{2,4,6,8\}$ and $\kappa_{c_{2}} \approx 1.80 \pm 0.02$ from the corresponding
curve using the data set $m=\{2,4,6\}$.  

In Fig.~\ref{M_diffSites} 
\begin{figure}[t]
\begin{center}
\includegraphics[angle=270,width=8cm]{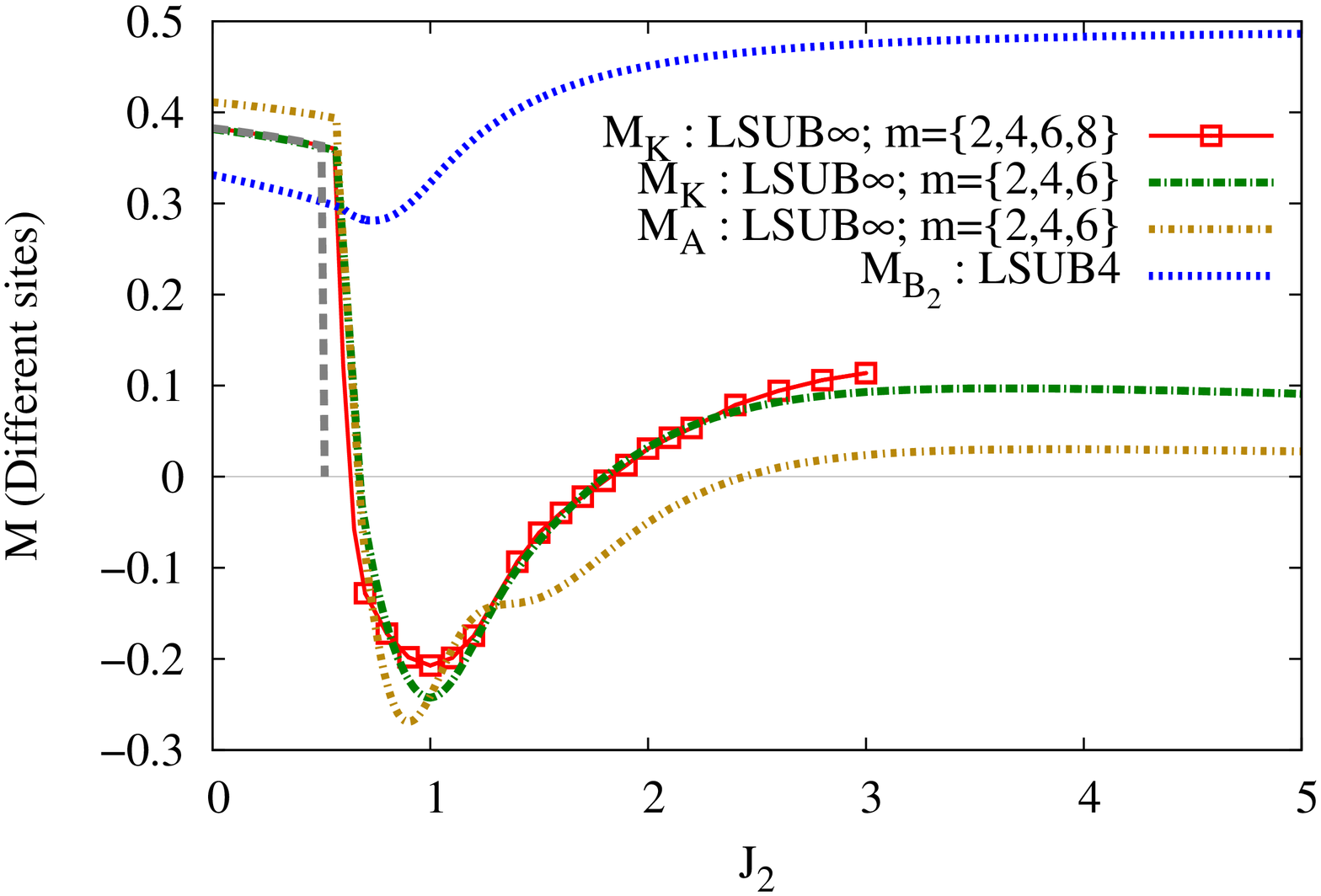}
\caption{
  (Color online) Various ground-state magnetic order parameters versus $J_{2}$ 
  for the spin-1/2 $J_{1}$--$J_{2}$
  HAF on the anisotropic kagome lattice of Eq.~(\ref{hamiltonian}) (with $J_{1} \equiv 1$), using
  the generic ferrimagnetic canted model state shown in Fig.~\ref{fig1}(b) as CCM model state, and
  with the canting angle $\phi=\phi_{{\rm LSUB}m}$ chosen to minimize the corresponding LSUB$m$ estimate
  for the energy, $E_{{\rm LSUB}m}(\phi)$, at each value of $J_2$.  Results are shown for the
  average local on-site magnetizations, $M_{{\rm A}}$ and $M_{{\rm B}_{2}}$, on the A sites and the
  B$_2$ sites respectively of Fig.~\ref{fig1}(b), as well as for their average value on the kagome
  lattice, $M_{K} = \frac{2}{3}M_{{\rm A}} + \frac{1}{3}M_{{\rm B}_{2}}$.  LSUB4 results are
  shown for $M_{{\rm B}_{2}}$, and
  extrapolated LSUB$\infty$ results are shown for $M_K$ using both data sets $m=\{2,4,6,8\}$ 
  and $m=\{2,4,6\}$, and for $M_{{\rm A}}$ using the data set $m=\{2,4,6\}$.  Extrapolated results 
  are calculated using Eq.~(\ref{M_extrapo_nu}) for the N\'{e}el$'$ state and
  Eq.~(\ref{M_extrapo_frustrated}) for the canted state, respectively; and  
  the (grey) dashed vertical line at $J_{2} = 0.515$ represents our best estimate for the termination point 
  above which collinear N\'{e}el$'$ order disappears, at 
  $J_{2} = J_{2}^{c_{1}} \equiv J_{2}^{{\rm LSUB}\infty}$, as discussed in the text.
  }
\label{M_diffSites}
\end{center}
\end{figure}
we also show, in various approximations, 
the separate average on-site magnetizations, $M_{{\rm A}}$ and $M_{{\rm B}_{2}}$, 
on the A sites and the B$_{2}$ sites respectively of Fig.~\ref{fig1}(b), as well as
their average value on the kagome lattice, $M_{K} = \frac{2}{3}M_{{\rm A}} + \frac{1}{3}M_{{\rm B}_{2}}$.  
We recall that the A sites are connected by two $J_{1}$ bonds and
two $J_{2}$ bonds, while the B$_2$ sites are connected by four $J_{1}$ bonds.  In particular,
we note that as $J_{2} \to \infty$ (with $J_{1} \equiv 1$), the model reduces to one of
independent linear HAF chains of alternating A$_1$ and A$_2$ sites.  In this limit
our extrapolated CCM result for the energy using the ferrimagnetic canted state
as model state (and see Fig.~\ref{E}) is $E/N_{K} \approx -0.2954J_{2}$.  Since in this limit
the B$_{2}$ sites become irrelevant, we may re-express the result in terms of the number, 
$N_{c} = \frac{2}{3}N_{K}$, of A sites that form the independent 1D chains in this limit, as
$E/N_{c} \approx -0.4431J_{2}$.  This compares extremely well
with the exact result for the 1D HAF, obtained from the Bethe ansatz solution,
$E/(N_{c}J_{2}) = \frac{1}{4} - \ln 2 \approx -0.443147$.

As $J_{2} \to \infty$ (with $J_{1} \equiv 1$) all of our CCM LSUB$m$ approximations give
$M_{{\rm B}_{2}} \to 0.5$, as expected.  The LSUB4 result for $M_{{\rm B}_{2}}$ 
shown in Fig.~\ref{M_diffSites} is typical
of the behavior of the entire LSUB$m$ set.  By contrast, in the same limit, the individual
LSUB$m$ approximations for $M_{{\rm A}}$ approach different constant (positive) values.  The 
extrapolated LSUB$\infty$ result for $M_{{\rm A}}$, shown in Fig.~\ref{M_diffSites}, however,
approaches a value very close to zero, again fully consistent with the exact behavior 
of 1D HAF chains.

Our results so far have indicated the presence of a paramagnetic state in the range
$\kappa_{c_{1}} < \kappa < \kappa_{c_{2}}$ between the quasiclassical states with N\'{e}el$'$
order (for $\kappa < \kappa_{c_{1}}$) and canted order (for $\kappa > \kappa_{c_{2}}$).  We
have seen too that as $\kappa \to \infty$ the model reduces consistently to the limit
of uncoupled isotropic HAF 1D chains.  There
remains still the question raised in Sec.~\ref{model_section} as to whether the canted
state illustrated in Fig.~\ref{fig1}(b) remains the stable GS phase all the way out
to $\kappa \to \infty$, where the canting angle $\phi \to 90^{\circ}$, or whether there might
exist a further phase transition at a value $\kappa = \kappa_{c_3} > \kappa_{c_2}$ to 
some other phase.  One such possibility is
the semi-striped state illustrated in Fig.~\ref{fig1}(c).  We have argued that such a phase
might be stabilized due to the quantum fluctuations
possibly lifting the infinite degeneracy, which exists in the classical counterpart between the  
orientation of the antiferromagnetically aligned spins on the A sites and the orientation
of the ferromagnetically aligned spins on the B$_{2}$ sites (that become decoupled 
from those on the A sites in this limit), by the order-by-disorder mechanism.  We consider
this possibility further in Sec.~\ref{discussions_conclusions}, where we also discuss
our results and compare them with those of others.

\section{Discussion and Conclusions}
\label{discussions_conclusions}

We have used the CCM to investigate the effects of quantum fluctuations on
the zero-temperature GS properties and phase diagram of the spin-1/2 $J_{1}$--$J_{2}$
HAF on the anisotropic kagome lattice of Eq.~(\ref{hamiltonian}), 
and as illustrated in Fig.~\ref{fig1}(b).  The system contains spatially anisotropic
NN exchange couplings on the kagome net, with coupling $J_{2} \equiv \kappa J_{1} > 0$ in one of the three
equivalent spatial directions of the lattice and coupling  $J_{1} \equiv 1$ along the
other two directions.  The model has only two classical GS phases.  For 
$\kappa < \kappa_{{\rm cl}} \equiv \frac{1}{2}$ the GS phase has collinear ferrimagnetic N\'{e}el$'$ order, in which
the $J_2$-chain spins on the A sites of the lattice are aligned in one direction
and the middle spins on the remaining B$_2$ sites are aligned in the opposite
direction.  At $\kappa = \kappa_{{\rm cl}} \equiv \frac{1}{2}$ the classical GS
configuration changes from the essentially unique collinear ferrimagnetic N\'{e}el$'$ 
state to an infinite ensemble of degenerate states.  These include a still infinite 
number of coplanar canted ferrimagnetic states that are expected to be selected from 
among the rest by thermal or quantum fluctuations to comprise
the stable GS phase for all $\kappa > \kappa_{{\rm cl}} \equiv \frac{1}{2}$.  

For a given value of $\kappa > \kappa_{{\rm cl}}$ all of these states are characterized by
a canting angle $\phi$ such that on each A$_1$A$_2$B$_2$ triangular plaquette of the kagome net
the two $J_2$-chain spins (i.e., those on the A$_1$ and A$_2$ sites) form angles ($\pi \pm \phi$) 
with respect to the middle spin (i.e., that on the B$_2$ site).  Clearly, the N\'{e}el$'$ 
state is just the special case with $\phi = 0$.  Each triangular plaquette 
in the ensemble of coplanar states also carries
a chirality variable, $\chi = \pm 1$, defined to be the direction (anticlockwise
or clockwise, respectively) in which the spins rotate as one traverses the plaquette
in the positive (anticlockwise) direction.  The different degenerate coplanar canted states
then correspond to different ways of assigning chiralities to the individual plaquettes.

In this paper we have used for the spin-1/2 model the generic ferrimagnetic canted model state shown in 
Fig.~\ref{fig1}(b) as our CCM model state, which is that member of the classically degenerate
ensemble in which all of the triangular plaquettes have the same (here positive) chirality.
This state thus corresponds to the $q=0$ state of the isotropic (i.e., when $J_{2}=1$)
kagome-lattice HAF, in which $\phi = \frac{\pi}{3}$ by symmetry. At each LSUB$m$ level
of approximation we have chosen the value of the canting angle $\phi$ that minimizes
the corresponding LSUB$m$ estimate for the GS energy.  We note again parenthetically that in
a previous recent CCM analysis\cite{Gotze:2011} of the isotropic kagome HAF in which 
both the $q=0$ state and the $\sqrt{3} \times \sqrt{3}$ state (that corresponds to 
that member of the classically degenerate ensemble in which the chiralities alternate, 
such that triangular plaquettes joined by a vertex have opposite values of $\chi$), it was 
found that for the extreme quantum case considered here, with $s=\frac{1}{2}$, 
the $q=0$ state is energetically favored over the $\sqrt{3} \times \sqrt{3}$ 
state, while for any $s>\frac{1}{2}$ the $\sqrt{3} \times \sqrt{3}$ state 
is selected over the $q=0$ state.  

We note too that previous CCM studies of many other strongly correlated
and highly frustrated models in quantum magnetism have shown that the calculated positions
of phase boundaries are rather insensitive to the choice of CCM model state where
several competing possibilities exist that lie close in energy to one another.
In the present case we have repeated the calculations performed here, for the
$q=0$ state as CCM model state, with the $\sqrt{3} \times \sqrt{3}$ state so chosen,
and have found that the effect on the resulting value of $\kappa_{c_2}$ is
basically within our stated error bars.

In a very interesting recent paper, Masuda {\it et al.}\cite{Masuda:2012} show that
a first-order phase transition, which has no counterpart in the isotropic 
case ($\kappa = 1$), occurs in the {\it classical} ($s \to \infty$) anisotropic
kagome model (with $\kappa > 1$) at a very low but finite temperature.  They conclude
that thermal fluctuations tend to favor, by the order-by-disorder mechanism,\cite{Vi:1977}
an incommensurate spiral phase from among the massively degenerate ensemble of classical
ground states for values of $\kappa > 1$.  This spiral state reduces to the 
$\sqrt{3} \times \sqrt{3}$ state in the isotropic limit $\kappa \to 1$.  Such a state,
however, must be extremely fragile to small perturbations in the Hamiltonian.  For
example, even an addition of the $J_{1}'$ bonds shown in Fig.~\ref{fig1}(a) with
an infinitesimal (positive) strength, acts to stabilize, for all values of 
$\kappa > \frac{1}{2}$, the classical canted state shown in the figure, which is
just the $q=0$ state used as our CCM model state.  The use of this spiral state as
CCM model state would again be very unlikely to alter our results for
$\kappa_{c_2}$ for reasons cited above.

We found here that the canting angle $\phi_{{\rm LSUB}m}$ that minimizes the GS energy 
$E_{{\rm LSUB}m}(\phi)$ for the spin-1/2 model, at a given LSUB$m$ level, becomes nonzero for values of the
anisotropy parameter $\kappa > \kappa^{{\rm LSUB}m}_{c_{1}}$, where $\kappa^{{\rm LSUB}m}_{c_{1}}$ is 
generally somewhat higher than the corresponding classical value 
$\kappa_{{\rm cl}} \equiv \frac{1}{2}$, as shown in Table~\ref{table_CritPt}.   
On the other hand these critical values converge to an extrapolated value
$\kappa_{c_1} \equiv \kappa_{c_1}^{{\rm LSUB}\infty}$ that is close to $\kappa_{{\rm cl}} \equiv \frac{1}{2}$.
Indeed our best estimate is $\kappa_{c_1} = 0.515 \pm 0.015$.  Although they
cannot entirely exclude the possibility of a small regime of canted phase
in a very narrow strip immediately above $\kappa_{c_1}$, our corresponding CCM
LSUB$m$ results for the magnetization shown in Figs.~\ref{M} and \ref{M_diffSites} give
compelling evidence, however, that the transition for the spin-1/2 model at 
$\kappa = \kappa_{c_1}$ from a GS phase with N\'{e}el$'$ order is {\it not} to one with canted order, 
as in the corresponding classical model at $\kappa = \kappa_{{\rm cl}} = 0.5$, but rather to a
paramagnetic state with no canted order.  Our evidence is that this paramagnetic GS
phase persists over the anisotropy range $\kappa_{c_1} < \kappa < \kappa_{c_2}$, before
the canted state becomes the stable GS phase for $\kappa > \kappa_{c_2}$. Our best estimate
for this upper critical point of the paramagnetic phase is $\kappa_{c_2} = 1.82 \pm 0.03$.  

Since the isotropic kagome-lattice point, $\kappa = 1$, is contained within the parameter
range $\kappa_{c_1} < \kappa < \kappa_{c_2}$ of this paramagnetic phase, the natural 
conclusion is that this phase of the anisotropic model shares the same order as the GS phase 
of the isotropic model.  As we discussed in Sec.~\ref{intro}, the isotropic spin-1/2 HAF on
the kagome lattice has been greatly studied in the past. The most direct results, namely those 
from the exact diagonalization (ED) of finite lattices,\cite{Lech:1997,Wald:1998,
Sindzingre:2009,Lauchli:2011,Nakano:2011a} seem to provide 
strong evidence for a spin-liquid GS phase.  This conclusion is supported by the results
of block-spin approaches\cite{Mila:1998,Mamb:2000} and by those from various other
studies too.\cite{Sach:1992,Leung:1993,Hast:2000,Wang:2006,Ran:2007,Herm:2008,Jiang:2008}

Very recent ED studies\cite{Lauchli:2011,Nakano:2011a} have examined
the GS energies and spin gaps (specifically between the GS singlet
level and the lowest-lying triplet level) of many isotropic kagome
clusters of sizes up to $N=42$.  In their study, for example, Nakano
and Sakai\cite{Nakano:2011a} further claim that, from the result of
their analysis of larger clusters, the isotropic HAF on the kagome
lattice is gapless, in contradiction with other recent DMRG
studies\cite{Yan:2011,Jiang:2012_kagome_spinLiquid,
  Depenbrock:2012_kagome_spinLiquid} that find it to be gapped.
Whereas earlier ED studies also attempted to resolve the spin-gap
issue, the data on clusters of sizes up to $N=36$ was
deemed\cite{Sindzingre:2009} to be insufficient to distinguish between
a gapless system and one with a very small gap.

On the other hand, conflicting results have been found by other
authors\cite{Mar:1991,Syr:2002,Nik:2003,Bud:2004,Sin:2007,Sin:2008,Eve:2010}
who have proposed various valence-bond solid states as the GS phase of
the isotropic HAF on the kagome lattice.  A detailed comparison of the exact
spectrum of a 36-site finite lattice sample of the isotropic kagome
HAF against the excitation spectra allowed by the symmetries of 
the various proposed valence-bond crystal states has, however, cast very strong doubts
on their validity as stable GS phases.\cite{Misg:2007}  Over the past year or so this
muddled and confused picture of the nature of the GS phase of the isotropic spin-1/2 HAF on
the kagome lattice has been seemingly resolved in favor of a topological
spin liquid.

In particular, as indicated in Sec.~\ref{intro}, the results of two independent and very
recent large-scale DMRG studies of the spin-1/2 isotropic HAF on the kagome 
lattice\cite{Jiang:2012_kagome_spinLiquid,Depenbrock:2012_kagome_spinLiquid} have provided
compelling {\it positive} evidence that this GS phase is a topological quantum spin liquid. The simplest such state
that preserves all symmetries is the $\mathbb{Z}_2$ spin liquid, and by explicitly
calculating the topological entanglement entropy both recent DMRG studies provide strong
positive evidence that the spin liquid state does indeed have $\mathbb{Z}_2$ topological order, with a finite spin (triplet) gap.\cite{Depenbrock:2012_kagome_spinLiquid}  

It has not been our aim here to investigate the order properties of the
paramagnetic GS phase of antiferromagnetically coupled $s=\frac{1}{2}$ spins on the 
infinite kagome lattice, but rather to investigate the stability of the phase as the
lattice is spatially distorted.  Clearly, however, 
it is natural to expect that over the entire range
$\kappa_{c_1} < \kappa < \kappa_{c_2}$ in which the paramagnetic phase persists (as
manifested here by a negative, and hence unphysical, value of the calculated local magnetic 
order parameter) it remains a spin liquid with the same topological order.

We are reticent to make specific claims of the direct relevance of our results to such 
real materials as volborthite.  Naturally we would like to be able to claim that the paramagnetic 
region $\kappa_{c_1} < \kappa < \kappa_{c_2}$ in which we have found that the classical ground 
states are unstable, has applicability to the spin glass phase observed experimentally in 
volborthite.  Indeed, if volborthite could certainly be described by the present anisotropic 
kagome model, then its value for $\kappa$ would fall within the paramagnetic region we have found, 
and such a claim might be justified.  Unfortunately, however, as we discussed in Sec.~\ref{intro}, 
the nature of the magnetic couplings in volborthite has recently been questioned.\cite{Janson:2010} 
Thus, it was pointed  out that the local environments of the two inequivalent types of Cu sites 
(that were previously used to justify the use of the present anisotropic model) 
differ in essential ways.  A DFT study\cite{Janson:2010} was then used to show 
that a better model of this material might be more akin to one involving coupled 
frustrated chains in which some of the NN bonds are actually 
{\it ferromagnetic} in nature.  

The spin-1/2 HAF on the spatially anisotropic kagome lattice has also been
studied by several other authors recently using a variety of
techniques.  These have included large-$N$ expansions of the
Sp($N$)-symmetric generalization of the actual SU(2) model,\cite{Apel:2007,Yavo:2007} a
block-spin perturbation approach to the trimerized kagome
lattice,\cite{Yavo:2007} various semiclassical calculations (appropriate to the limit of
large spin quantum number $s$) that include (a) studying an effective chirality 
Hamiltonian derived from a low-temperature classical nonlinear spin-wave 
expansion,\cite{Wang:2007} and (b) keeping terms of order $1/n$ in the 
large-$n$ limit of the O($n$) generalization of the classical O(3) 
model together with a high-temperature expansion,\cite{Wang:2007} field-theoretical
techniques appropriate to quantum critical systems in one dimension
(and which are hence appropriate here for the case $\kappa \equiv J_{2}/J_{1} \gg 1$
of weakly coupled chains),\cite{Kagome_Schn:2008} and a renormalization-group
analysis in the same quasi-1D limit but now also in the presence of a
Dzyaloshinskii-Moriya interaction.\cite{Zyuzin:2012}

Since many of these calculations employ perturbation theories of one kind or another
in some ``artificial'' small parameter, direct comparison is difficult.  Thus, for example,
in the large-$N$ Sp($N$) expansion, the effective smallness parameter is 
$\alpha \equiv n_{b}/N$, where $n_b$ is the number of bosons on each site.  While in the
physical SU(2) model (which corresponds to the case $N=1$) we have $\alpha = 2s$, the 
comparison in the large-$N$ limit actually studied is lost.  Yavors'kii {\it et al.}\cite{Yavo:2007} 
argue that the value of $\alpha$ that corresponds to the $s=\frac{1}{2}$ under study
must be somewhat less than 0.5.  Similarly, in
the block-spin perturbation approach of Yavors'kii {\it et al.}\cite{Yavo:2007} the assumption
is made that the kagome lattice is trimerized such that the spins on the downward-pointing triangles, say,
are strongly coupled whereas the couplings on the bonds of the upward-pointing triangles are
weaker by a factor $\gamma$.  The approximate GS phases of this trimerized model are then
studied in different regimes of the anisotropy parameter $\kappa$ in a perturbation 
expansion with respect to $\gamma$, while the physical model corresponds to
the case $\gamma = 1$.  

Complementary to such essentially perturbative studies have
been various more controlled analyses of the quasi-1D limit of the 
model.\cite{Kagome_Schn:2008,Zyuzin:2012}  The latter studies, by their nature
of focussing on the large-anisotropy ($\kappa \gg 1$) limit, also lose sight
of the intermediate paramagnetic (spin-liquid) phase that has been the focus 
of the present study.  For example, Zyuzin {\it et al.}\cite{Zyuzin:2012} explicitly
state that they do not find a spin-liquid ground state in any regime that they study.

As we have seen, much of the previous work on the spin-1/2 HAF on the spatially
anisotropic kagome lattice has approached the quantum limit only very indirectly, either from the
classical side or in such slave particle approaches as the Schwinger boson technique
applied in the large-$N$ Sp($N$) approach.  The only direct $s=\frac{1}{2}$ approach
seems to be a small-scale ED study of up to $N_{K}=24$ spins (in a $4 \times 2$ unit cell arrangement)
with periodic boundary conditions.\cite{Wang:2007}  The numerical evidence from the ED study
seems to indicate very clearly that for values of the anisotropy parameter
$\kappa < 0.5$ the GS phase has nonzero total spin.
Indeed, its value for the few finite-size lattices studied is precisely what is expected 
for the classical collinear N\'{e}el$'$ ferrimagnetic state, namely $S_{{\rm tot}}=\frac{1}{3}N_{K}s$, 
thereby agreeing fully with our own findings.  By contrast, the numerical evidence for
values $\kappa > 0.5$ is far less clear.  While the evidence seems to be
that for all lattice sizes up to 24 sites the GS phase is a spin singlet for all values
$\kappa > 0.5$, there is a clear tendency for a state of nonzero spin at the $\Gamma$
point to drop in energy on moving away from the isotropic point $\kappa = 1$, perhaps
indicating a tendency to develop a net moment again.  Nevertheless, the evidence
from such small-scale ED studies seems to leave completely open the nature
of the GS phase for $\kappa > 0.5$.

The two semiclassical approaches of Wang {\it et al.}\cite{Wang:2007} also
concur that for $\kappa < 0.5$ the GS phase is the collinear N\'{e}el$'$ ferrimagnetic state.
For the case $\kappa >1$ both approaches also indicate a canted ferrimagnetic state of the 
classical type, but where the infinitely-degenerate manifold of coplanar states is
lifted by the order-by-disorder mechanism, which now seems to favor the so-called
{\it chirality stripe} state as the GS phase, in which all spins on the interstitial B$_2$ sites are 
ferromagnetically aligned, and where pairs of triangles on the 
kagome lattice that share either an A$_1$ or A$_2$ vertex have the same value of the 
chirality parameter $\chi$, while pairs sharing a B$_2$ vertex have opposite values of
$\chi$.  (We note parenthetically that this state is the only other state, apart
from the $q=0$ state of Fig.~\ref{fig1}(b), that is allowed by 
the chirality constraints in the general case $\kappa \neq 1$ and 
in which all of the B$_2$-spins are aligned parallel to 
one another.)  By contrast, in the intermediate regime $0.5 < \kappa <1$, the semiclassical
models do not give clear indications of which GS ordering is favored.  Thus, while the 
$\sqrt{3} \times \sqrt{3}$ order seems to be favored in the spin-wave expansion for the
isotropic case $\kappa = 1$, the comparable analysis for the $\kappa < 1$ case seems to
depend very sensitively on both the choice of unphysical parameters and on the number
of chirality-chirality couplings included in the analysis.  By contrast, the large-$n$
saddle-point solution of the O($n$) generalization of the classical O(3) 
model seems to favor the $q=0$ GS ordering of spins for the case $0.5 < \kappa <1$.

The large-$N$ Sp($N$) expansion analysis of the model\cite{Apel:2007,Yavo:2007} indicates
that the actual spin-1/2 HAF on the spatially anisotropic kagome lattice has a GS phase
with collinear N\'{e}el$'$ ferrimagnetic order for small values of the anisotropy parameter
$\kappa$, which gives way at larger values of $\kappa$ to an incommensurate (spiral) GS
phase with no LRO, in general agreement with our findings.  In turn, as $\kappa$ is
increased further this GS phase then gives way to a phase in which the chains are
completely decoupled, while the interstitial spins (i.e., those on B$_2$ sites)
show some short-range spin-spin correlations.  For reasons already
noted above this Sp($N$) analysis is unable to give quantitative estimates for the corresponding
two critical values of $\kappa$ for the actual spin-1/2 SU(2) model.

Finally, we note that the canted ferrimagnetic phase that we have found to be
the stable GS phase for $\kappa > \kappa_{c_2}$, after the disappearance of the
paramagnetic (spin-liquid) phase is not likely to remain the
GS phase for sufficiently large values of $\kappa$, as we have already noted in
Sec.~\ref{model_section}, since in the limit $\kappa \to \infty$ the canting angle
$\phi \to \frac{\pi}{2}$, and the chain spins become perpendicular to the 
interstitial spins.  Since in this limit the relative orientation of the chain spins
and the interstitial spins becomes irrelevant, and since quantum fluctuations generally
prefer collinear spin configurations in such a degenerate situation, 
we thus expect a third phase transition at a value
$\kappa = \kappa_{c_3}$ to a phase that eventually becomes the decoupled 1D HAF chain phase
in the asymptotic limit $\kappa \to \infty$.  The precise nature of this fourth phase 
is by no means settled.  

We have alluded in Sec.~\ref{model_section} to one such candidate
being the collinear ferrimagnetic semi-striped phase shown in
Fig.~\ref{fig1}(c).  We are currently investigating whether this
phase might become energetically favored at sufficiently large values of $\kappa$.  
We intend to report on this in a separate future paper.

An alternative candidate state for the large-$\kappa$ phase has been suggested by
Yavors'kii {\it et al.}\cite{Yavo:2007} from their block-spin trimerized version of the model.  
In the subsequent small-$\gamma$ perturbative limit they find a tentative ground state in the
large-anisotropy case ($\kappa \gg 1$) that is a collinearly ordered antiferromagnet 
in which the interstitial (B$_2$) spins are N\'{e}el-ordered and the spins on the A-chains
form singlet dimers (i.e., the spins on each A$_1$-A$_2$ NN pair of sites on, say, each
downward-pointing triangle on the kagome lattice form a spin-singlet state).  

Yet another analysis of the large-$\kappa$ limit, by Schnyder {\it et al.},\cite{Kagome_Schn:2008} 
suggests that all of the spins order with a (generally noncoplanar) configuration in which 
the interstitial spins on B$_2$ sites and the chain spins on A-sites each separately form
predominantly coplanar spirals with a wave vector $(q,0)$, but with a reduced [$O(1/\kappa)$]
static moment on the $J_2$-coupled chains.  These authors find that the chain spins are 
weakly canted out of the plane, with the [$O(1/\kappa^{2})$] normal components being
ordered in an antiferromagnetic fashion.  While their analysis could not determine $q$ reliably,
it is expected that $q \ll 1$ and, indeed, the authors suggest that $q=0$ is a real
possibility, in which case the state becomes coplanar.  Nevertheless, even this state differs
from both the chirality stripe state considered by Wang {\it et al.}\cite{Wang:2007} and
the dimerized state considered by Yavors'kii {\it et al.}\cite{Yavo:2007}, although
there are some similarities with each. 

It is clear that the large-$\kappa$ limit of the spin-1/2 HAF on the anisotropic kagome lattice
is far from settled.  Our own work presented here has mainly been concerned 
to investigate the stability with respect to anisotropy $\kappa$ of the spin-liquid state that
has convincingly been found in recent work to be the stable GS phase of the isotropic 
($\kappa = 1$) model.  Nevertheless, we hope to return in the future to the quite separate question 
of whether or not there is a further 
transition at some value $\kappa = \kappa_{c_3} > \kappa_{c_2}$ of the anisotropy 
parameter, from the canted state discussed here to some other
state with or without collinear order.

\section*{ACKNOWLEDGMENTS}
We thank the University of Minnesota Supercomputing Institute for
Digital Simulation and Advanced Computation for the grant of
supercomputing facilities, on which we relied heavily for the
numerical calculations reported here.

\end{document}